\documentclass[]{elsarticle}

\usepackage{graphicx}
\usepackage[framemethod=TikZ]{mdframed}
\usepackage{amsmath}
\usepackage[flushleft]{threeparttable}
\usepackage[caption=false]{subfig}
\usepackage{hyperref}

\mdfsetup {outerlinewidth=1pt,
outerlinecolor=gray!15,
middlelinewidth=1pt,
middlelinecolor=white,
backgroundcolor=gray!15,
roundcorner=1pt,
skipabove=.4\baselineskip}

\journal{Journal on System and Software}

\usepackage{xspace}
\usepackage{xcolor}

\newcommand{\yw}{YW\xspace}
\newcommand{\tdd}{TDD\xspace}

\newcommand{\mra}{MRA\xspace}
\newcommand{\bsk}{BSK\xspace}
\newcommand{\ssh}{SSH\xspace}
\newcommand{\gol}{GOL\xspace}

\newcommand{\qlty}{QLTY\xspace}
\newcommand{\pro}{PROD\xspace}
\newcommand{\test}{TEST\xspace}
\newcommand{\seq}{SEQ\xspace}
\newcommand{\gra}{GRA\xspace}
\newcommand{\uni}{UNI\xspace}
\newcommand{\re}{REF\xspace}
\newcommand{\mut}{MUT\xspace}

\newcommand{\ie}{\textit{i.e.,}\xspace}
\newcommand{\eg}{\textit{e.g.,}\xspace}
\newcommand{\etc}{\textit{etc.}\xspace}
\newcommand{\etal}{\textit{et al.}\xspace}

\begin{document}

\begin{frontmatter}

\title{Studying Test-driven Development and its Retainment Over a Six-month Time Span}

\author[label1]{Maria Teresa Baldassarre}
\ead{mariateresa.baldassarre@uniba.it}

\author[label1]{Danilo Caivano}
\ead{danilo.caivano@uniba.it}

\author[label2]{Davide Fucci}
\ead{davide.fucci@bth.se}

\author[label3]{Natalia Juristo}
\ead{natalia@fi.upm.es}

\author[label1]{Simone Romano\corref{cor1}}
\ead{simone.romano@uniba.it}

\author[label4]{Giuseppe Scanniello}
\ead{giuseppe.scanniello@unibas.it}

\author[label5,label6]{Burak Turhan}
\ead{burak.turhan@monash.edu}

\address[label1]{University of Bari, Bari, Italy}
\address[label2]{Blekinge Institute of Technology, Karlskrona, Sweden}
\address[label3]{Universidad Politecnica de Madrid, Madrid, Spain}
\address[label4]{University of Basilicata, Potenza, Italy}
\address[label5]{Monash University, Melbourne, Australia}
\address[label6]{University of Oulu, Oulu, Finland}
\cortext[cor1]{Corresponding Author}





\begin{abstract}
Test-Driven Development (\tdd) is an approach to agile software development, which is claimed to boost both external quality of software products and developers' productivity. The results about the claimed effects of \tdd are inconclusive, therefore researchers have recommended taking a longitudinal perspective when studying \tdd---\ie studying \tdd over a time span. By following such a recommendation, we investigated: \textit{(i)} the retainment of \tdd over a time span of about six months, as well as \textit{(ii)} the effects of \tdd. To pursue our two-fold objective, we conducted a quantitative longitudinal cohort study with 30 novice developers (\ie third-year undergraduate students in Computer Science).  We found that \tdd is retained by developers over a time span of about six months. As for the effects of \tdd, we observed that this development approach affects neither the external quality of software products nor developers' productivity. However, we observed that the participants applying \tdd produced significantly more tests, with a higher fault-detection capability, than those using a non-\tdd approach.

\end{abstract}


\end{frontmatter}

\section{Introduction}
Test-Driven Development (\tdd)~\cite{Bec03,Ast03} is a cyclic development approach where unit tests drive the incremental development of small pieces of functionality~\cite{Erdogmus:2010}. Each development cycle starts with the writing of unit tests for an unimplemented piece of functionality. A cycle ends when  unit tests pass as well as the existing regression test suite. An important role in the process underlying \tdd is played by refactoring. It allows a \tdd practitioner to improve the internal structure of the code, as well as its design, while preserving the external behavior of the code thanks to the safety net the existing regression test suite provides~\cite{Ast03}. The end of a cycle allows a \tdd practitioner to tackle a new piece of functionality, not yet implemented, so starting a new development cycle~\cite{Bec03,Ast03}.
Advocates of \tdd recommend ending a development cycle in few minutes (five or ten minutes~\cite{Jeffries:2007}) and keeping the rhythm as uniform as possible over time~\cite{Bec03,Erdogmus:2010}. The order with which unit tests interpose within the process underlying \tdd---\ie the writing of a test precedes the one of the corresponding production code---is known as \textit{test-first sequencing} (or also \textit{test-first dynamic})~\cite{Fucci:2017}. It is worth noting that test-first sequencing refers to just one central aspect of \tdd~\cite{Karac:2018}. That is, it does not capture the full nature of \tdd~\cite{Fucci:2017}. Other central aspects that characterize the development process underlying \tdd are: \textit{granularity}, \textit{uniformity}, and \textit{refactoring effort}~\cite{Fucci:2017}. Granularity refers to the duration of the development cycles, while uniformity reflects how constant their duration is over time~\cite{Fucci:2017}. Finally, refactoring effort captures how much refactoring a \tdd practitioner performs.

It is claimed that \tdd leads to higher-quality products in terms of both external (\ie functional) and internal quality, while increasing developers' productivity~\cite{Bec03}. These claimed benefits have encouraged some software companies to adopt \tdd, while others are considering its adoption~\cite{Tosun:2017}. 
\tdd has been assessed from a quantitative point of view (\eg \cite{Fucci:2016,Erdogmus:2005}) and according to a qualitative perspective (\eg \cite{Romano:2016,Scanniello:2016}). A number of primary studies, like experiments or case studies, have been conducted on \tdd (\eg \cite{Fucci:2016,Erdogmus:2005,George:2004,Bhat:2006,Nagappan:2008}). Their results, gathered and combined in a number of secondary studies (\eg \cite{Karac:2018,Bissi:2016,FTJ15,TLD06,MMP14,RM13}), do not fully support the claimed benefits of~\tdd. Therefore, some researchers have recommended taking a longitudinal perspective when investigating such a development approach (\eg~\cite{FTJ15,MMP14,SMT10,MH07})---\ie studying \tdd over a time span. Nevertheless, only a few studies have taken such a perspective~(\eg~\cite{Lat14}).



Longitudinal studies\footnote{There are three major types of longitudinal studies: \textit{(i)} repeated cross-sectional studies; \textit{(ii)} prospective studies (including cohort studies); and \textit{(ii)} retrospective studies~\cite{Caruana:2015}.} employ continuous or repeated measures to follow particular individuals over a time span of weeks, months, or even years~\cite{Caruana:2015}. 
In this paper, we present a study on \tdd that takes a longitudinal perspective. In particular, we conducted a longitudinal cohort study in which our \textit{cohort} consisted of 30 novice software developers of homogeneous experience who attended the same training regarding agile software development, including \tdd. Thanks to that cohort, we collected separate measurements of the same constructs (\ie external quality,  developers' productivity, number of tests written, fault-detection capability of tests written, test-first sequencing, granularity, uniformity, and refactoring effort) about six months apart with the goal of understanding how well \tdd can be applied over time, giving an indication of its \textit{retainment}. 
To have a term of comparison, we contrasted \tdd with a non-\tdd approach (\eg iterative test-last, big-bang testing, or no testing at all), namely the approach that developers would normally follow. We refer to the non-\tdd approach as \textit{Your Way} (\ie~YW). Our results indicate that novice developers retain \tdd over a time span of about six months. Moreover, although we did not find any improvement, due to \tdd, in the external quality of software products and developers' productivity, we observed that \tdd allows creating larger test suites with a higher fault-detection~capability. 

This paper extends the one by Fucci~\etal~\cite{Fucci:2018:ESEM} as follows:
\begin{itemize}
    \item We investigated the retainment of \tdd with respect to four aspects that characterize the process underlying \tdd: test-first sequencing, granularity, uniformity, and refactoring effort.
    \item Since Fucci \etal had found that \tdd leads developers to write more tests, we studied whether writing more tests implies that the fault-detection capability of those tests is actually better. This was to strengthen the conclusions from Fucci \etal's study. It is worth mentioning that we studied both effect and retainment of \tdd with respect to fault-detection capability of written tests.
    \item We extended the inferential statistics by applying a second statistical model. This allowed us to mitigate as much as possible threats to the conclusion validity of the results shown in the paper of Fucci \etal.
    \item We extended the discussion of results. This extension is directly related to the several extensions and improvements of our data analysis which has allowed strengthening our conclusions.
    \item We improved the discussion of both related work and threats to validity.
\end{itemize}


\textbf{Paper structure.} In Section~\ref{sec:background}, we outline work related to ours. We present our study in Section~\ref{sec:study}. The obtained results are presented and discussed in Section~\ref{sec:results} and Section~\ref{sec:discussion}, respectively. Final remarks conclude the paper.

\section{Related Work}\label{sec:background}

In this section, we describe the kinds of longitudinal studies researchers have conducted in the context of software engineering. Moreover, we show the results from secondary studies on the claimed effects of \tdd---\ie better external quality and increased developer productivity---as well as those from long-term investigations on \tdd.

\subsection{Longitudinal studies in Software Engineering}

The goal of a longitudinal study is to investigate \textit{``how certain conditions change over time''}~\cite{Yin09}.
Therefore, the data collection happens over a time span and can require the researchers to be co-located with the case and context in which the phenomenon of interest takes place. In the context of software engineering, longitudinal studies are often associated with the case study methodology.
For example, McLeod~\etal~\cite{MMD11} spent several hundred hours, over a time span of two years, at the case company.
Here, the researcher attended meetings, observed and interviewed stakeholders to characterize software development as an emergent process.

In other cases, longitudinal studies are employed to observe the impact of a potentially disrupting event, such as the introduction of a new development practice.
This scenario is similar to interrupted time series in quasi-experimental designs~\cite{CCS02} in which, due to the lack of experimental manipulation, a specific event is used to identify the experimental groups.
For example, Li \etal~\cite{LMD10} studied the differences between Scrum and a waterfall-like approach to software development in a small company.
The authors collected data from the development team for three years.
In the first 17 months, the team used a waterfall-like approach before switching to Scrum for the following 20 months.
The management decision to introduce Scrum represented the event allowing researchers to perform a \textit{before-after} comparison of defects density and productivity.
Such an extended time span avoided a biased comparison between an established process and an immature one.
Salo and Abrahamson~\cite{SA05} followed the introduction of Software Process Improvement (SPI) techniques in the workflow of five agile projects over a time span of 18 months.
They recorded the output of retrospective meetings, interviewed developers, and collected metrics from SPI tools.
Vanhanen \etal~\cite{VLM07} assessed the impact of introducing pair programming over a time span of two years with data collected through a survey with the~developers.

A third kind of longitudinal study in software engineering retrospectively covers an extended time span by analyzing archival data.
Harter \etal~\cite{HKS12} analyzed the type of defects identified over time by the progressive introduction of SPI (Software Process Improvement) techniques in a firm and its subsequent CMMI (Capability Maturity Model Integration) improvements over a time span of 20 years.
Given the availability of a large amount of versioned and timestamped data, longitudinal archival studies are usually performed in conjunction with software repositories mining studies. For example, Borges \etal~\cite{BHV16} studied the growth over time in popularity of more than 2,000 GitHub repositories by identifying useful patterns for the maintainers.

\subsection{Evidence regarding \tdd} \label{sec:relatedwork}
The effects of \tdd on several outcomes, including the ones of interest for this study---\ie functional quality and productivity---is the topic of several empirical studies, summarized in Systematic Literature Reviews (SLRs) and meta-analyses (\eg~\cite{Bissi:2016,TLD06,MMP14,RM13}). The SLR by
Turhan~\etal~\cite{TLD06} includes 32 primary studies (\eg controlled experiments and case studies) published from 2000 to 2009. The gathered evidence shows a moderate effect in favor of \tdd on functional quality while the evidence about productivity is inconclusive.\footnote{It means that the results do not lead to a firm conclusion.} Bissi \etal~\cite{Bissi:2016} conducted an SLR that includes 27 primary studies published between 1999 and 2014. The results show an improvement of functional quality due to \tdd while, as for productivity, they are inconclusive. Rafique and Misic~\cite{RM13} conducted a meta-analysis of 25 controlled experiments published in between 2000 and 2011. The authors observed a small effect in favor of TDD on functional quality while the results on productivity are inconclusive. Finally, Munir \etal~\cite{MMP14} in their SLR classifies 41 primary studies published from 2000 to 2011 into four categories based on high/low rigor and high/low relevance. They found that in each category different conclusions could be drawn for both functional quality and productivity. This implies that, when looking at these studies as a unique set, the results are inconclusive. The authors concluded that more long-term studies are needed to better understand the effects of TDD.

An example of long-term investigation is the one by Marchenko \etal~\cite{MAI09}.
The authors conducted a three-year-long case study about the use of \tdd at Nokia-Siemens Network.
They observed and interviewed eight participants (one Scrum master, one product owner, and six developers) and then ran qualitative data analyses.
The participants perceived \tdd as important for the improvement of their code from a structural and functional perspective.  Moreover, productivity increased due to the team improved confidence with the code base.
The results show that \tdd was not suitable for bug fixing, especially when bugs are difficult to reproduce (\eg when a specific environment setup is needed) or for quick experimentation due to the extra effort required for testing.
The authors also reported some concerns regarding the lack of a solid architecture when applying \tdd.

Beller \etal~\cite{BGP17} executed a long-term study \textit{in-the-wild} covering 594 open-source projects over the course of 2.5 years. They found that only 16 developers use \tdd more than 20\% of the time when making changes to their source code.
Moreover, \tdd was used in only 12\% of the projects claiming to do so, and for the majority by experienced developers.

Borle \etal~\cite{Borle:2017} conducted a retrospective analysis of (Java) projects, hosted on GitHub, that adopted \tdd to some extent. The authors built sets of \tdd projects that differed one another based on the extent to which \tdd was adopted within these projects. The sets of \tdd projects were then compared to control sets so as to determine whether \tdd had a significant impact on the following characteristics: average commit velocity, number of bug-fixing commits, number of issues, usage of continuous integration, and number of pull requests. The results did not suggest any significant impact of \tdd on the above-mentioned characteristics.

Latorre~\cite{Lat14} studied the capability of 30 professional software developers of different seniority levels (junior, intermediate, and expert) to develop a complex software system by using \tdd.
The study targeted the \textit{learnability} of \tdd since the participants did not know that technique before participating in the study. The longitudinal one-month study started after giving the developers, proficient in Java and unit testing, a tutorial on \tdd. After only a short practice session, the participants were able to correctly apply \tdd (\eg following the prescribed steps).
They followed the \tdd cycle between 80\% and 90\% of the time, but initially, their performance depended on experience.
The seniors needed only few iterations, whereas intermediates and juniors needed more time to reach a high level of conformance to \tdd.
Experience had an impact on performance---when using \tdd, only the experts were able to be as productive as they were when applying a traditional development methodology (measured during the initial development of the system).
According to the junior participants, refactoring and design decision hindered their performance.
Finally, experience did not have an impact on long-term functional quality. The results show that all participants delivered functionally correct software regardless of their seniority.
Latorre~\cite{Lat14} also provides initial evidence on the retainment of \tdd. Six months after the study investigating the learnability of \tdd, three developers, among those who had previously participated in that study, were asked to implement a new functionality. The results from this preliminary investigation suggest that developers retain \tdd in terms of developers' performance and conformance to~\tdd.

Although the above-mentioned studies~\cite{Lat14,MAI09,BGP17,Borle:2017} have taken a longitudinal perspective when studying \tdd, none of them has mainly focused, as our longitudinal cohort study, on the retainment of \tdd---although Latorre's study~\cite{Lat14} provides initial evidence on the retainment of \tdd, the main goal of this study was the learnability of \tdd.

\begin{figure}[t]
      \includegraphics[width=\linewidth]{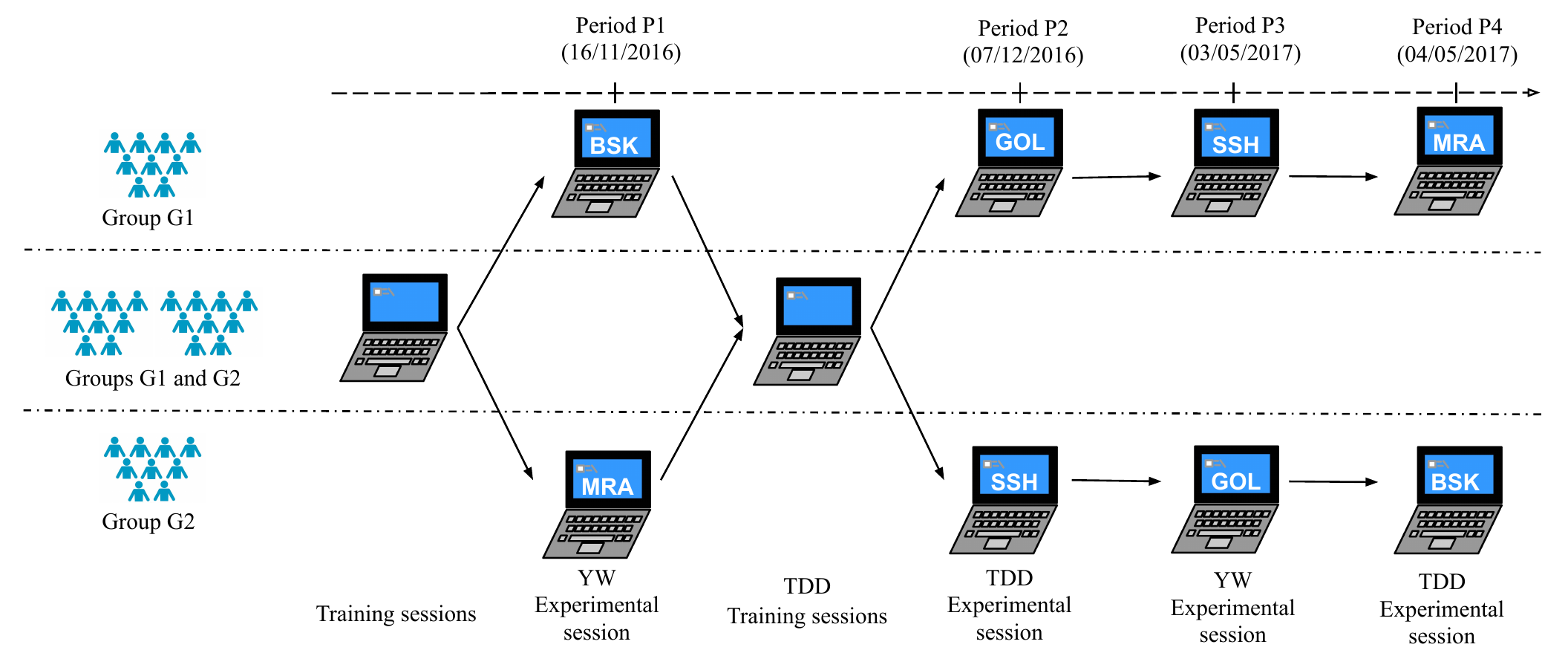}
\caption{Study summary.}\label{fig:summary}
\end{figure}

\section{The Longitudinal Cohort Study}\label{sec:study}
In Figure~\ref{fig:summary}, we summarize the design of our longitudinal cohort study.
The study participants were third-year undergraduate students in Computer Science. They were divided into two groups G1 and G2 and were asked to take part in four experimental sessions. In any experimental session, each participant had to perform a development task by following either the \yw approach or the \tdd approach. Before the first experimental session, all the participants had practiced unit testing, iterative test-last development, and big-bang testing thanks to training sessions.

In the first experimental session, which was held in the period P1---a period is the time during which a treatment is applied~\cite{Vegas:2016}--- the participants  had to use the \yw approach to perform development tasks. The two groups were asked to accomplish different development tasks---\ie G1 dealt with an experimental object (BSK, \ie Bowling Score Keeper) while G2 dealt with another one (MRA, Mars  Rover  API). We provide further details on the experimental objects in Section~\ref{sec:ExpMaterial}.

The participants in G1 and G2 learned and practiced (together) \tdd between the first experimental session and the second one (held in the period P2). That is to say no one in the first experimental session knew \tdd. It is easy to grasp that the first experimental session was introduced to have a baseline when the participants were not knowledgeable on \tdd yet.

In the second experimental session, all the participants used \tdd to perform the development tasks. The tasks assigned to G1 and G2 were different. In particular, G1 worked on GOL (Game Of Life), while G2 on SSH (SpreadSHeet).

After about six months (over two subsequent semesters of the same academic year, 2016-2017), we asked the participants to take part in the third and fourth experimental sessions, which were held in the periods P3 and P4 respectively. In particular, in the third experimental session, all the participants followed the \yw approach when performing the development tasks, while in the fourth experimental session, they followed the \tdd approach. We introduced the last two periods (\ie P3 and P4) some months after the first two periods to assess whether \tdd was~retained. It is worth mentioning that the tasks we asked the groups G1 and G2 to perform  were different both in P3 (SSH and GOL, respectively) and P4 (MRA and BSK, respectively). 

To plan and execute our cohort study, we followed the recommendations by Juristo and Moreno~\cite{Juristo:2001}, and Wohlin \etal~\cite{Wohlin:2012}. To report the planning and the execution of that study, we took into account the guidelines by Jedlitschka \etal~\cite{Jedlitschka2008}.

\subsection{Research Questions}

In our longitudinal cohort study, we have investigated the following main Research Questions (RQs):
\begin{description}
\item[\textbf{RQ1.}] Do novice software developers retain \tdd?

RQ1 was defined to study whether the retainment of \tdd affects the application of \yw, as well as the application of \tdd, over approximately six months. We studied the retainment of \tdd with respect to the following constructs: \textit{(i)}~external quality of the implemented solutions, \textit{(ii)}~developers' productivity, \textit{(iii)}~number of tests written,  \textit{(iv)}~fault-detection capability of tests written, \textit{(v)}~test-first sequencing, \textit{(vi)}~granularity, \textit{(vii)}~uniformity, and \textit{(viii)}~refactoring effort. We considered external quality of the implemented solutions and developers' productivity because \tdd is claimed to improve external quality of the implemented solutions and increase developers' productivity~\cite{Bec03}. We focused on the number of tests because past research has shown that \tdd results in more tests~\cite{Erdogmus:2005}. However, having a test suite with more tests could not imply an increased fault-detection capability of that suite; therefore, we also studied the fault-detection capability of tests written. Finally, we considered test-first sequencing, granularity, uniformity, and refactoring effort because these are the four dimensions that characterize the development process underlying \tdd~\cite{Fucci:2017}. It is worth recalling that the study of these four dimensions, together with the one on the fault-detection capability of tests written, is a new contribution with respect to our previous paper~\cite{Fucci:2016}.

\item[\textbf{RQ2.}] Are there differences between \tdd and \yw?

We defined RQ2 to study if there are differences, due to \tdd, in the \textit{(i)}~external quality of the implemented solutions, \textit{(ii)}~developers' productivity, \textit{(iii)}~number of tests written, and \textit{(iv)}~fault-detection capability of tests written. The reasons behind the study of these constructs have been explained above. We did not consider the four dimensions that characterize the development process underlying \tdd because these dimensions do not characterize \yw. With respect to the paper by Fucci \etal~\cite{Fucci:2016}, we also investigated whether or not \tdd leads to a higher fault-detection capability of tests written---\ie this is another new contribution of this paper.
\end{description}



\subsection{Experimental Units}
The participants were third-year undergraduate students in Computer Science at the University of Bari (Italy). We sampled them by convenience among the students who attended the \textit{Integration and Testing} course (first semester of the academic year 2016/2017). The program of this course included the following topics: unit testing, integration testing, SOLID principles, refactoring, big-bang testing, iterative test-last development, and \tdd.
During the course, the students participated in both face-to-face lessons and laboratory sessions. The students practiced unit testing, big-bang testing, iterative test-last development, and \tdd through laboratory sessions and some homework was assigned too. Java was the programming language of the course, while JUnit and Eclipse were the testing framework and the Integrated Development Environment~(IDE), respectively. Among the 53 students of the Integration and Testing course, 39 decided to take part in the study. The first two experimental sessions of our study were held during the Integration and Testing course.

Some students of the Integration and Testing course then attended the \textit{Software Quality} course (second semester of the academic year 2016/2017). The program of this course included the following topics: software quality (\ie internal, external, and in-use); ISO standards for software quality; software quality assessment, monitoring, and improvement; supporting tools for quality management (\eg SonarQube); and process control. The students enrolled in the Software Quality course were 45, 30 of them took part in the third and fourth experimental sessions. These 30 students had previously attended (and passed) the Integration and Testing course and had participated to the first two experimental sessions. This is to say that the intersection of the students who attended both courses (\ie Integration and Testing and then Software Quality) and participated in the longitudinal cohort study (\ie in any of the fourth experimental sessions) were 30---two females and 28 males.

\begin{figure}[t]
\subfloat[]{\label{fig:javaExperience:peers}
      \includegraphics[width=0.85\linewidth]{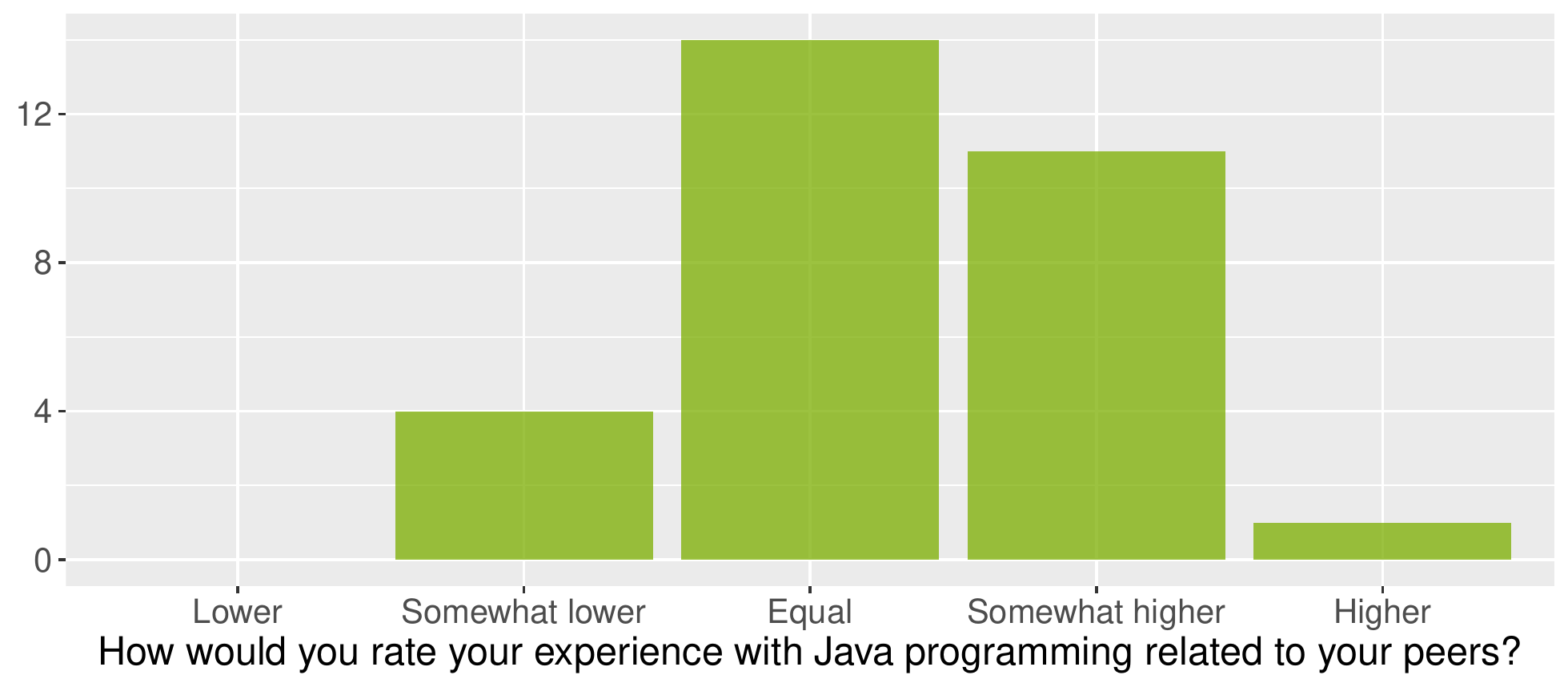}}

\subfloat[]{\label{fig:javaExperience:general}
      \includegraphics[width=0.85\linewidth]{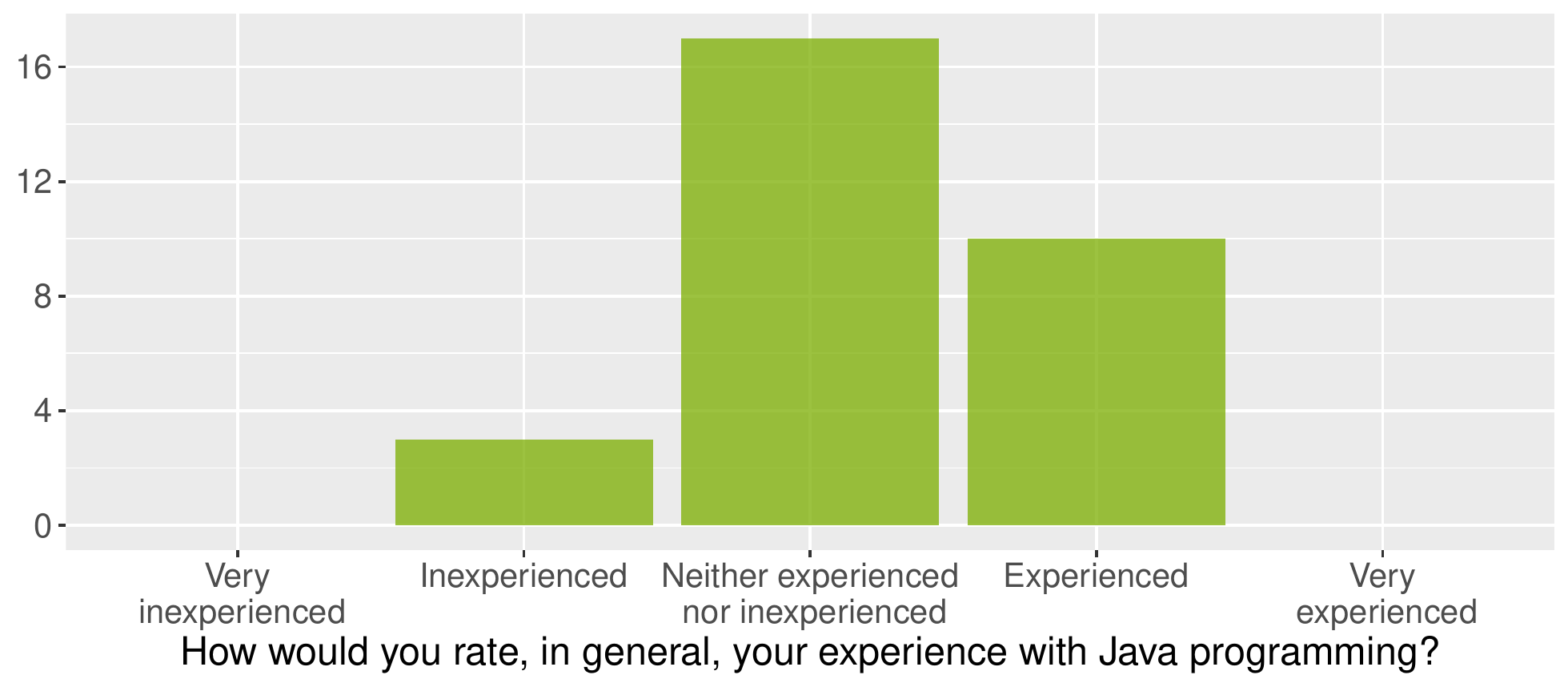}}

\caption{Barplots on the participants' experience with Java programming related to (a) their peers and (b) in general.}\label{fig:javaExperience}
\end{figure}

Before participating in the study, the students did not have a notion of \tdd since their university curricula did not include courses on \tdd. The participants had passed the exams of the following courses: \textit{Procedural Programming}, \textit{Object Oriented Programming}, \textit{Software Engineering}, and \textit{Databases}. Thanks to these courses, the participants had gained experience in C and Java programming. As shown in Figure~\ref{fig:javaExperience:peers}, most of the participants rated their experience as equal to or somewhat higher than that of their peers. Only four participants believed to be somewhat less expert than their peers. Figure~\ref{fig:javaExperience:general} shows how the participants generally rated their experience with Java programming. Most of them stated to be neither experienced nor inexperienced. Only three participants judged themselves as inexperienced with Java programming. Summing up, the participants' characteristics can be considered homogeneous.



To encourage the students to participate in the study, we informed them that they would be rewarded with a bonus in the final mark of the Integration and Testing course. The students also knew that their participation would not affect their final mark (except for the bonus mentioned just before) and that the gathered data would be used only for research purposes. It is worth mentioning that students could not be paid for their participation in the study because this is forbidden in Italy (while rewarding them with a bonus in their final mark is allowed). Participation was voluntary in the sense that the students were not coerced to participate. All these choices were made to have motivated participants even if we were conscious that it could represent a threat to the internal validity of the results (see Section~\ref{sec:internalValidity}). We also informed the students that the collected data would have been treated confidentially and shared anonymously.

\subsection{Experimental Materials} \label{sec:ExpMaterial}
The experimental objects were four code katas (\ie programming
exercises used to practice a programming language or a development approach like \tdd). A description of these code katas follows:
\begin{itemize}
    \item \textbf{Bowling Score Keeper (BSK).} The goal of this kata is to develop an API for calculating the score of a bowling game made up of ten frames (plus potential bonus throws). The API allows: adding frames and bonus throws to a bowling game; identifying if a frame is a spare or a strike; and computing the score of a single frame as well as the score of a bowling~game.
    \item \textbf{Mars Rover API (MRA).} This kata aims at developing an API for moving a rover on a planet. The planet is represented as a grid of cells, which can contain obstacles that the rover cannot go through. The rover moves thanks to a sting made up of basic commands (\ie moving forward/backward and turning left/right). When the rover encounters an obstacle, it records that obstacle.
    \item \textbf{SpreadSHeet (SSH).} The goal of this kata is to develop an API for a basic spreadsheet. This API allows setting the content of a spreadsheet's cell and evaluating its content. A cell can contain strings, integers, references to other cells, and formulas (\eg string concatenations or arithmetic operations among integers).
    \item \textbf{Game Of Life (GOL).} This kata aims at developing an API for Conway's game of life. This game takes place on a square grid of cells. Each cell has two possible states: alive or dead. The state of a cell evolves according to four rules. The API allows: initializing the grid; and determining the next state of a cell as well as the next state of the grid.
\end{itemize}
For each code kata (or experimental object,  from here onwards), the experimental material included a template project for the Eclipse IDE, which contained stubs of the expected API signatures and an example JUnit test class. The code katas could be broken down into several features to implement; however, the description of the code katas was \textit{coarser-grained}. That is, each code kata was presented as a whole without explicitly identifying the features to be implemented---in contrast to a \textit{finer-grained} description of code katas in which the features to be implemented are described separately, thus they are explicitly identified (\eg each feature is numbered)~\cite{Karac:2019}. To assess the features the participants implemented, the experimental material also included acceptance test suites. In particular, there was an acceptance test suite for each feature being implemented. The participants were not provided with these acceptance test suite because their purpose was the assessment of the implemented features. So, the acceptance test suites were used to quantify the external quality of the solutions implemented by the participants as well as their productivity.


Our decision to adopt code katas is because their use is common in empirical studies on \tdd (\eg~\cite{Fucci:2017,Tosun:2017,Fucci:2016,Erdogmus:2005,Karac:2019,Dieste:2017}). Furthermore, this allowed us to use existing experimental materials (\eg from the studies by Fucci~\etal~\cite{Fucci:2016} and Dieste \etal~\cite{Dieste:2017}). 
BSK and MRA have been also implied as experimental objects in several empirical studies (\eg~\cite{Fucci:2017,Tosun:2017,Fucci:2016,Karac:2019,Dieste:2017}).
Whereas for SSH and GOL, we created the experimental materials (\ie description of code katas, template projects, and acceptance test suites).

To gather some information on the participants (\eg  gender,  self-reported experience with Java programming, \etc), we defined an on-line pre-questionnaire we shared with the participants through Google Forms. We also created on-line post-questionnaires (by using Google Forms) to gather feedback after the participants had dealt with the code katas.

\subsection{Tasks}
The participants were asked to carry out four implementation tasks, one for each experimental object (see Figure~\ref{fig:summary}). To this end, each participant received the features to be implemented and the template project of a code kata, thus he/she implemented the features by filling the provided template project. No graphical user interface was required to implement the features.



\subsection{Independent Variables}\label{sec:IV}

To carry our an implementation task, the participants were asked to follow either  \tdd  or  \yw  (\ie the approach they preferred, excluding \tdd). Therefore, \textbf{Approach} is the main independent variable (or also main  or manipulated factor) of our study. This variable is nominal and assumes two possible values: \textit{\tdd} and \textit{\yw}. Since our study is longitudinal---\ie we collected data over time---we had a second main independent variable we named \textbf{Period}. It is a nominal variable that represents the period during which each treatment (\ie \tdd or \yw) was applied. Therefore, this variable can assume the following values: \textit{P1}, \textit{P2}, \textit{P3}, and \textit{P4}. It is worth recalling that P1 and P3 correspond to the application of \yw, while P2 and P4 correspond to that of \tdd (see Figure~\ref{fig:summary} for details).

\subsection{Dependent Variables}\label{sec:dep:vars}
To quantify external quality of the implemented solutions, developers productivity, number of tests written, we used the following dependent variables: \textbf{\qlty}, \textbf{\pro}, and \textbf{\test}. We chose these dependent variables because they have been used in other empirical studies on \tdd~(\eg~\cite{Fucci:2017,Tosun:2017,Fucci:2016,Erdogmus:2005}).

The variable \qlty measures the external quality of the solution to a code kata a participant implemented. This variable is defined as follows (\eg~\cite{Fucci:2016}):
\begin{equation}\label{eq:1}
\qlty=\frac{\sum_{i=1}^{\#TF} \qlty_i}{\#TF} * 100
\end{equation}
where \#TF is the number of features a participant tackled, while QLTY\textsubscript{i} is the external quality of $i$-th feature.
A feature was tackled if at least one assert in the acceptance test suite (for that feature) passed. \#TF is formally defined as~follows:
\begin{equation}\label{eq:2}
\#TF=\sum_{i=1}^n \bigg \{ \begin{array}{rl}
1 & \#ASSERT_i(PASS)>0 \\
0 & \textstyle otherwise \\
\end{array}
\end{equation}
As for QLTY\textsubscript{i}, it is computed as the number of asserts passed for the i-th feature divided by the total number of asserts for that feature:
\begin{equation}\label{eq:3}
QLTY_i=\frac{\#ASSERT_i(PASS)}{\#ASSERT_i(ALL)}
\end{equation}
\qlty assumes values in between 0 and 100. A value close to 0 means that the quality of the implemented solution is low, while a value close to 100 indicates high quality of the implemented solution.


As for the variable \pro, it measures the productivity of a participant when carrying out the implementation task. \pro is defined as follows (\eg ~\cite{Tosun:2017}):
\begin{equation}\label{eq:4}
\pro=\frac{\#ASSERT(PASS)}{\#ASSERT(ALL)}*100
\end{equation}
where \#ASSERT(PASS) is the number of asserts passed in the acceptance test suites, while \#ASSERT(ALL) is the total number of asserts in the acceptance test suites. PROD assumes values in between 0 and 100, where a value close to 0 means low productivity, while a value close to 100 means high productivity.

The variable \test quantifies the number of unit tests a participant wrote. It is defined as the number of asserts in the test suite written by a participant when tackling the implementation task (\eg~\cite{Fucci:2016}). \test assumes (integer) values in between 0 and $\infty$. A high value is desirable.

To quantify fault-detection capability of tests written, we leveraged mutation testing~\cite{Jorgensen:2013}. Given a program, mutation testing consists of automatically seeding artificial faults (\ie mutation faults) to generate mutants, each of which represents a faulty version of that program. Later, the test suite of the program is run against the mutants to determine the extent to which the test suite is capable of killing the generated mutants (\ie detecting the corresponding mutation faults). For each solution implemented by a participant, we seeded mutation faults into his/her production code  (\ie we did not seed any fault in the test code) so generating mutants. To this end, we used the \textit{Major} mutation framework~\cite{Just:2014}. We opted for this framework because it is robust~\cite{Papadakis:2018}, publicly available~\cite{Just:2014}, and has been adopted in previous work (\eg~\cite{Papadakis:2018,Just:2014:FSE}). We applied the following mutation operators\footnote{They alter a program by systematically applying a rule (\eg they replace the $+$ arithmetic operator with the $-$ one).} to generate mutants: AOR, LOR, COR, ROR, ORU, LVR, and STD. A description of these operators is available in Table~\ref{tab:operators}. This set of mutation operators is the same as that Papadakis \etal~\cite{Papadakis:2018} used in their empirical investigation on the relationship between mutation and real faults. We ran the test suite the participant had written against the generated mutants so computing the MUTation score (\textbf{\mut}), namely the dependent variable we used to estimate fault-detection capability of tests written. MUT is computed as follows (\eg ~\cite{Jorgensen:2013}):
\begin{equation}\label{eq:7}
\mut=\frac{\#MUTANTS(KILLED)}{\#MUTANTS(ALL)}*100
\end{equation}
where \#MUTANTS(KILLED) is the number of mutants the test suite killed, while \#MUTANTS(ALL) is the total number of generated mutants. \mut assumes values in the interval [0,100]. The greater \mut, the better it is. In particular, it has been proven that \mut values close to 100 imply a higher fault-detection capability as compared with \mut values close to 0~\cite{Papadakis:2018}. This is why we leveraged mutation testing to estimate the fault-detection capability of written tests.

\begin{table}[t]
\centering
\caption{Description of the mutation operators.}\label{tab:operators}
\resizebox{\linewidth}{!}{
\begin{tabular}{ll}
\hline\noalign{\smallskip}
Mutation operator & Description \\ \noalign{\smallskip}\hline\noalign{\smallskip}
AOR (Arithmetic Operator
Replacement) & Replaces an arithmetic operator (\eg $+$) with another one (\eg $-$) \\
LOR (Logical Operator Replacement) & Replaces a logical operator (\eg $\&$) with another one (\eg $|$)\\
COR (Conditional Operator Replacement) & Replaces a conditional operator (\eg $\&\&$) with another one (\eg $||$)\\
ROR (Relational Operator Replacement) & Replaces a relational operator (\eg $>$) with another one (\eg $>=$) \\
ORU (Operator Replacement Unary) & Replaces a unary operator (\eg $++$) with another one (\eg $--$) \\
LVR (Literal Value Replacement) & Replaces a literal value (\eg 0) with a default value (\eg 1)\\
STD (STatement Deletion) & Deletes a single statement (\eg a return statement) \\ \noalign{\smallskip}\hline
\end{tabular}
}
\end{table}

Besides the above-mentioned constructs, we investigated four constructs dealing with the development process underlying \tdd, namely: test-first sequencing, granularity, uniformity, and refactoring effort. To quantify these constructs, we broke down the development process of participants applying \tdd into small cycles as done by Fucci \etal~\cite{Fucci:2017}. A cycle consists of a sequence of elementary actions and ends with a successful regression testing (\ie the regression test suite does not highlight regressions). Thanks to the heuristics devised by Kou \etal~\cite{Kou:2009}, it is possible to determine the type of each cycle (\eg test-first or refactoring). In Table~\ref{tab:heuristics}, we report the heuristics we exploited to determine the type of the cycles when the participants applied \tdd. The considered heuristics are implemented in \textit{Besouro}~\cite{Becker:2015}.

\begin{table}[!t]
\centering
\caption{Heuristics implemented in Besouro~\cite{Becker:2015} to determine the type of cycles (the description of these heuristics is taken from Kou \etal' paper~\cite{Kou:2009}).}
\resizebox{\linewidth}{!}{
\begin{tabular}{ll}
\hline\noalign{\smallskip}
Cycle type & Sequence of actions  \\ \noalign{\smallskip}\hline\noalign{\smallskip}
Test-first & Test creation $\rightarrow$ Test compilation error $\rightarrow$ Code editing $\rightarrow$Test failure $\rightarrow$ Code editing $\rightarrow$ Test pass \\
                               & Test creation $\rightarrow$ Test compilation error $\rightarrow$ Code editing $\rightarrow$ Test pass                                                       \\
                               & Test creation $\rightarrow$ Code editing $\rightarrow$ Test failure $\rightarrow$ Code editing $\rightarrow$ Test pass                                      \\
                               & Test creation $\rightarrow$ Code editing $\rightarrow$ Test pass                                                                                             \\ \noalign{\smallskip}
Refactoring   & Test editing (file size changes $\pm$ 100 bytes) $\rightarrow$ Test pass                                                                                                                           \\
                               & Code editing (number of methods, or statements decrease) $\rightarrow$ Test pass                                                                                                                           \\
                               & Test editing AND Code editing $\rightarrow$ Test pass                                                                                                      \\ \noalign{\smallskip}
Test addition & Test creation $\rightarrow$ Test pass                                                                                                                          \\
                               & Test creation $\rightarrow$ Test failure $\rightarrow$ Test editing $\rightarrow$ Test pass                                                                 \\ \noalign{\smallskip}

Production & Code editing (number of methods unchanged, statements increase) $\rightarrow$ Test pass \\
								& Code editing (number of methods increase, statements increase) $\rightarrow$ Test pass \\
								& Code editing (size increases) $\rightarrow$ Test pass  \\ \noalign{\smallskip}
Test-last & Code editing $\rightarrow$ Test creation $\rightarrow$ Test editing $\rightarrow$ Test pass                                                                  \\
                               & Code editing $\rightarrow$ Test creation $\rightarrow$ Test editing $\rightarrow$ Test failure $\rightarrow$ Code editing $\rightarrow$ Test pass  \\ \noalign{\smallskip}
Unknown & None of the above  $\rightarrow$ Test pass \\ \noalign{\smallskip}\hline
\end{tabular}
}
\label{tab:heuristics}
\end{table}

The test-first sequencing construct indicates the prevalence of test-first sequencing within development processes underlying \tdd. We quantified this construct by means of the \textbf{\seq} dependent variable, which is defined as follows~\cite{Fucci:2017}:
\begin{equation}\label{eq:5}
\seq=\frac{\#CYCLES(TEST\operatorname{-}FIRST)}{\#CYCLES(ALL)}*100
\end{equation}
where \#CYCLES(TEST-FIRST) is the number of cycles classified as test-first by applying the heuristics in Table~\ref{tab:heuristics} when a participant followed \tdd. \#CYCLES(ALL) is, instead, the total number of cycles for that participant. The \seq variable assumes values in between 0 and 100. The higher the value for this variable, the higher the amount of test-first cycles when a participant applied~\tdd. 

Granularity refers to the extent to which the development process underlying \tdd is fine-grained (or coarse-grained). To estimate this construct, we used the \textbf{\gra} dependent variable. It is computed as the median duration (expressed in minutes) of the development cycles a participant carried out~\cite{Fucci:2017}. This variable ranges between 0 and $\infty$. A low \gra value indicates that a participant mostly carried out short cycles---\ie his/her development process was fine-grained. On the other hand, a high \gra value indicates that a participant tended to carry out long cycles---\ie his/her development process was coarse-grained. The use of median to compute \gra, rather than mean, allows reducing the impact of~outliers~\cite{Fucci:2017}.

Uniformity indicates how uniform the development process underlying \tdd is. This construct is quantified by means of the \textbf{\uni} variable, which is computed as the Median Absolute Deviation (MAD) of the cycle duration. This variable ranges between 0 and $\infty$. The lower the \uni value, the more uniform the cycles carried out by a participant were. A \uni value equal to 0 means that the cycles had mostly the same duration. The use of MAD to compute \gra, rather than standard deviation, allow reducing the sensitivity to outliers~\cite{Fucci:2017}.

Refactoring effort indicates the prevalence of refactoring within the development process underlying \tdd. We used the variable \textbf{\re} to estimate the refactoring effort construct. It is computed as follows~\cite{Fucci:2017}:
\begin{equation}\label{eq:6}
\re=\frac{\#CYCLES(REFACTORING)}{\#CYCLES(ALL)}*100
\end{equation}
where \#CYCLES(REFACTORING) is the number of cycles classified as refactoring (by using the heuristics in Table~\ref{tab:heuristics}) when a participant followed \tdd. The \re variable assumes values in between 0 and 100. The higher the value for this variable, the higher the refactoring effort of a participant when he/she applied~\tdd.

Since test-first sequencing, granularity, uniformity, and refactoring effort characterize the development process underlying \tdd, we took into account these constructs only when the participants applied the \tdd approach (\ie within periods P2 and~P4).

\subsection{Hypotheses}

We formulated and investigated the following parameterized null hypotheses:
\begin{description}
\item[\textbf{HN1\textsubscript{X}.}] There is no statistically significant effect of Period with respect to X (\ie \qlty, \pro, \test, \mut, \seq, \gra, \uni, or \re).
\item[\textbf{HN2\textsubscript{X}.}] There is no statistically significant effect of Approach with respect to X (\ie \qlty, \pro, \test, or \mut).
\end{description}
The alternative hypotheses were two-tailed (\ie whatever the independent variable was, we did not consider the direction of its effect). We defined HN1\textsubscript{X} to study RQ1, while HN2\textsubscript{X} to study RQ2.

\subsection{Study Design}
In Table~\ref{tab:design} (and Figure~\ref{fig:summary}), we summarize the design of our cohort study. We randomly split the participants into G1 and G2, each of which had 15 participants. Whatever the group was, the participants experimented each treatment (\ie \tdd or \yw) twice. In particular, both groups experimented: \textit{(i)}~\yw in the first period (\ie P1); \textit{(ii)}~\tdd in the second period (\ie P2); \textit{(iii)}~\yw in third period (\ie P3); and \textit{(iv)}~\tdd in the last period (\ie P4). Accordingly, the design of our study is \textit{repeated measures} (or \textit{within-subjects}).
In each period, the participants in G1 and G2 dealt with different experimental objects. For example, in P1, the participants in G1 dealt first with \bsk, while those in G2 dealt first with \mra. At the end of the study, the participants had tackled each experimental object only once.

\begin{table}[t]
\centering
\caption{Summary of the study design.}
\label{tab:design}

\scriptsize
\begin{tabular}{llllll}
\cline{3-6}\noalign{\smallskip}
& & \multicolumn{4}{l}{{Period}} \\ 
                   &  &  {P1 (16/11/2016)} & {P2 (07/12/2016) }& {P3 (03/05/2017)} &{ P4 (04/05/2017)} \\ \noalign{\smallskip}\hline\noalign{\smallskip}
Group & G1 & YW,  BSK      & \tdd, GOL      & YW, SSH       & \tdd, MRA      \\
& G2 & YW,  MRA      & \tdd, SSH      & YW, GOL       & \tdd, BSK      \\ \noalign{\smallskip}\hline
\end{tabular}
\end{table}

\subsection{Procedure}
The Integration and Testing course---\ie the course in which the experimental sessions corresponding to P1 and P2 took place---started in October 2016. A mentioned before, we gathered some demographic information on the participant through an on-line pre-questionnaire at the beginning of that course.

The first experimental session (corresponding to the period P1) took place on November 16th, 2016. In particular, we administered the participants with the \yw treatment. Between the beginning of the course and P1, the participants had never dealt with \tdd. On the other hand, they had knowledge of unit testing, iterative test-last development, and big-bang testing. This is because the participants had taken part in two training sessions and carried out some homework (see Figure~\ref{fig:summary}).

The first application of the \tdd treatment took place on December 7th, 2016 (\ie P2). The participant learned \tdd between P1 and P2. They had taken part in three training sessions on \tdd and had completed some homework by using this development practice. Given the previous considerations, we can exclude that the knowledge of \tdd had affected the application of the YW treatment~in~P1.

The participants applied the YW treatment again on May 3rd, 2017 (\ie P3), while the second application of the \tdd treatment happened on May 4th, 2017 (\ie P4).
Periods P3 and P4 took place in the Software Quality course.
From P2 to P3 passed about six months---over this span of time, the participants followed the same university curricula courses. Although we asked the participants to use \tdd, they knew \tdd in P3. Therefore, we cannot exclude that the knowledge of \tdd had affected the YW treatment in P3 somehow---\ie if the \tdd retainment had affected the (second) application of YW or not. On the other hand, we assessed the retainment of \tdd by asking the participants to use \tdd (once again) in P4.

The execution of the longitudinal cohort study as additional teaching activities of the Integration and Testing and Software Quality courses somewhat imposed the time span we considered in that study. This is to say that a larger time span could not represent an feasible alternative in our case since the Software Quality course represented the only alternative to catch the largest number of students who had previously attended the Integration and Testing course. Moreover, the considered time span allowed us to counteract the following problems that are typical in longitudinal cohort studies: participants sometimes drop out, while others could loose the motivation to participate. It is worth recalling that the considered time span of about six months to study the retainment of \tdd is similar to the one by Latorre~\cite{Lat14} (\ie the only study that has somehow investigated the retainment of \tdd, see Section~\ref{sec:relatedwork}).

The implementation tasks were executed, under controlled conditions, in a laboratory at the University of Bari. In each period, the participants in G1 and G2 were assigned to the PCs in the laboratory---this laboratory was also used for the training sessions. When assigning the participants to the PCs, we alternated a participant in G1 with a participant in G2 to avoid that participants dealing with the same experimental object were close to one another. Such an arrangement aimed to avoid interactions among the participants. We also monitored them during the execution of the tasks.

The PCs in the laboratory were all equipped with the same hardware and software. On these PCs, we had installed Eclipse with the Besouro plugin~\cite{Becker:2015}. Each participant received the description of code kata to be implemented, while, on the PC, he/she found the template project (for Eclipse) corresponding to that code kata. To carry out the implementation tasks, the participants had to use Java, JUnit, and Eclipse. At the beginning of a task, the participants launched Besouro within Eclipse, which started gathering data on their development process. These data allowed us to determine the type of development process of the participants who applied the \tdd approach. At the end of each task,  participants uploaded their implemented solutions on GitHub and then filled out a post-questionnaire to gather feedback on the executed task.

\subsection{Analysis Procedure}\label{sec:analysis}
We analyzed the gathered data according to the following procedure:
\begin{enumerate}
    \item \textbf{Descriptive Statistics and Exploratory Analyses.} We computed descriptive statistics (\ie mean, median and standard deviation), to summarize the distributions of the dependent variable values. To graphically summarize these distributions, we also  boxplots.
    \item \textbf{Inferential Statistics.} We used the Linear Mixed Model (LMM) analysis method to test the defined null hypotheses. Such a method is appropriate for the analysis of data from longitudinal studies~\cite{Verbeke:2010}. LMMs are an extension of linear models containing both fixed and random effects. As for HN1\textsubscript{X}, we built, for the dependent variable X (\eg QLTY), the following~LMM:
    \begin{equation}
        LMM1_X = X \sim Period + Group + Period:Group + (1|Participant)
    \end{equation}
    where \textbf{Period} (\ie the main independent variable), \textbf{Group}---it assumes G1 or G2 as value---, and their interaction (\ie \textbf{Period:Group}) are the fixed effects. LMM1\textsubscript{X} also includes a random effect, namely \textbf{Participant} (it identifies each participant, \eg \textit{01} is the first participant). Modeling the participants with a random effect is customary in software engineering experiments~\cite{Vegas:2016}. When building LMM1\textsubscript{X}, we took into account Group because, according to the study design, it also represents the sequence (\ie the order in which the treatments are applied in combination with the experimental objects). The sequence effect should be analyzed in repeated-measures designs (like ours)~\cite{Vegas:2016}.

    It is worth recalling that the periods P1 and P3 correspond to the application of the \yw treatment, while P2 and P4 correspond to that of the \tdd treatment. This implies that if LMM1\textsubscript{X} does not indicate a statistically significant effect of Period then there is not even a statistically significant effect of Approach. Accordingly, HN2\textsubscript{X} cannot be rejected. On the other hand, if there is a statistically significant effect of Period, the effect of Approach can be statistically significant. Therefore, only when LMM1\textsubscript{X} allowed us to reject HN1\textsubscript{X}, we built a second LMM that included Approach (instead of Period) to test HN2\textsubscript{X}:
    \begin{equation}
        LMM2_X = X \sim Approach + Group + Approach:Group + (1|Participant)
    \end{equation}
    If LMM2\textsubscript{X} revealed a statistically significant effect of Approach, we rejected HN2\textsubscript{X}. The use of LMM2\textsubscript{X} is new with respect to the paper by Fucci \etal~\cite{Fucci:2018:ESEM}.

    LMMs have two assumptions that must be met: \textit{(i)} the model residuals must be normally distributed; and \textit{(ii)} their mean must be zero~\cite{Vegas:2016}. In case LMM assumptions are not met, transforming the dependent variable values is an option (\eg log-transformation)~\cite{Vegas:2016}. To check the normality of the residuals, we used the Shapiro-Wilk test (Shapiro test, from here~onwards)~\cite{Shapiro:1965}.

    Whatever the test of statistical significance was, we set the $\alpha$ value at 0.05---\ie we accepted a probability of 5\% of committing Type-I error (this is customary in software engineering experiments).
\end{enumerate}

\section{Results}\label{sec:results}
In the following of this section, we first present the results from the descriptive statistics and exploratory analyses and then we provide results from the inferential statistics. 

\subsection{Descriptive Statistics and Exploratory Analyses}
In Table~\ref{tab:stats}, we report, for each dependent variable, mean, median, and Standard Deviation (SD) grouped by Period and Approach.

\begin{table}[!pt]
\caption{Descriptive statistics for each dependent variable grouped by Period and Approach.}
\label{tab:stats}
\resizebox{\linewidth}{!}{
\begin{tabular}{llllllll} \hline\noalign{\smallskip}
Variable   & Statistic & \multicolumn{4}{l}{Period (Approach)}  & \multicolumn{2}{l}{Approach} \\
   & & P1 (\yw) & P2 (\tdd) & P3 (\yw) & P4 (\tdd)  & \yw & \tdd \\ \noalign{\smallskip}\hline\noalign{\smallskip}
\qlty &  Mean &  59.39 & 63.1 & 63.05 & 58.53  &   61.22 & 60.81 \\
&  Median     &    76.76 & 69.72 & 71.28 & 74.76  &   72.97 &  71.99 \\
&  SD            & 37.85 & 31.98 & 30.73 & 34.58  & 34.23 & 33.11 \\ \noalign{\smallskip}\hline\noalign{\smallskip}

\pro &  Mean         & 34.11 & 32.47 & 30.99 & 37.96  & 32.55 & 35.22 \\
 &  Median         & 27.52 & 29.06 & 27.9 & 42.85  & 27.9 & 34.88 \\
&  SD & 32.18 & 29.03 & 28.97 & 29.19 &  30.4 & 29 \\ \noalign{\smallskip}\hline\noalign{\smallskip}

{\test} &  Mean         & 4.93  &  7.83  &  7.93   & 10.1  & 6.43  & 8.96 \\
&  Median & 4 & 6.5 & 5 & 8.5 & 5 & 7 \\
&  SD & 4.05 & 5.52 & 7.51 & 7.24 & 6.17 & 6.48 \\ \noalign{\smallskip}\hline\noalign{\smallskip}

{\mut} &  Mean & 31.98 & 32.07 & 31.99 & 48.52 & 31.99 & 40.29  \\
& Median  & 24.1 & 37.32 & 35.43 & 48.5 & 34.25 & 40.91 \\
& SD  & 30.97 & 20.78 & 23.65 & 25.18 & 27.32 & 24.34 \\
\noalign{\smallskip}\hline\noalign{\smallskip}

{\seq} & Mean & - & 27.91 & - & 22.3 & - & 25.1 \\
&  Median & - & 19.21 & - & 19.09 & - & 19.09 \\
&  SD & - & 25.73 & - & 20.27 & - & 23.1 \\ \noalign{\smallskip}\hline\noalign{\smallskip}

{\gra} &  Mean & - & 10.29 & - & 4.68 & - & 7.49 \\
&  Median & - & 4.26 & - & 2.5 & - & 3.03 \\
&  SD & - & 16.33 & - & 6.47 & - & 12.62 \\ \noalign{\smallskip}\hline\noalign{\smallskip}

{\uni} &  Mean & - & 5.76 & - & 2.82 & - & 4.29 \\
&  Median & - & 3.22 & - & 1.74 & - & 2.33 \\
&  SD & - & 6.5 & - & 4 & - & 5.54 \\ \noalign{\smallskip}\hline\noalign{\smallskip}

{\re} & Mean & - & 22.89 & - & 23.69 & - & 23.29 \\
&  Median & - & 18.61 & - & 25.36 & - & 21.98 \\
&  SD & - & 17.4 & - & 14.22 & - & 15.73 \\ \noalign{\smallskip}\hline
\end{tabular}
}
\end{table}

\subsubsection{\qlty---External Quality of Implemented Solutions}
\begin{figure}[t]
      \includegraphics[width=0.9\linewidth]{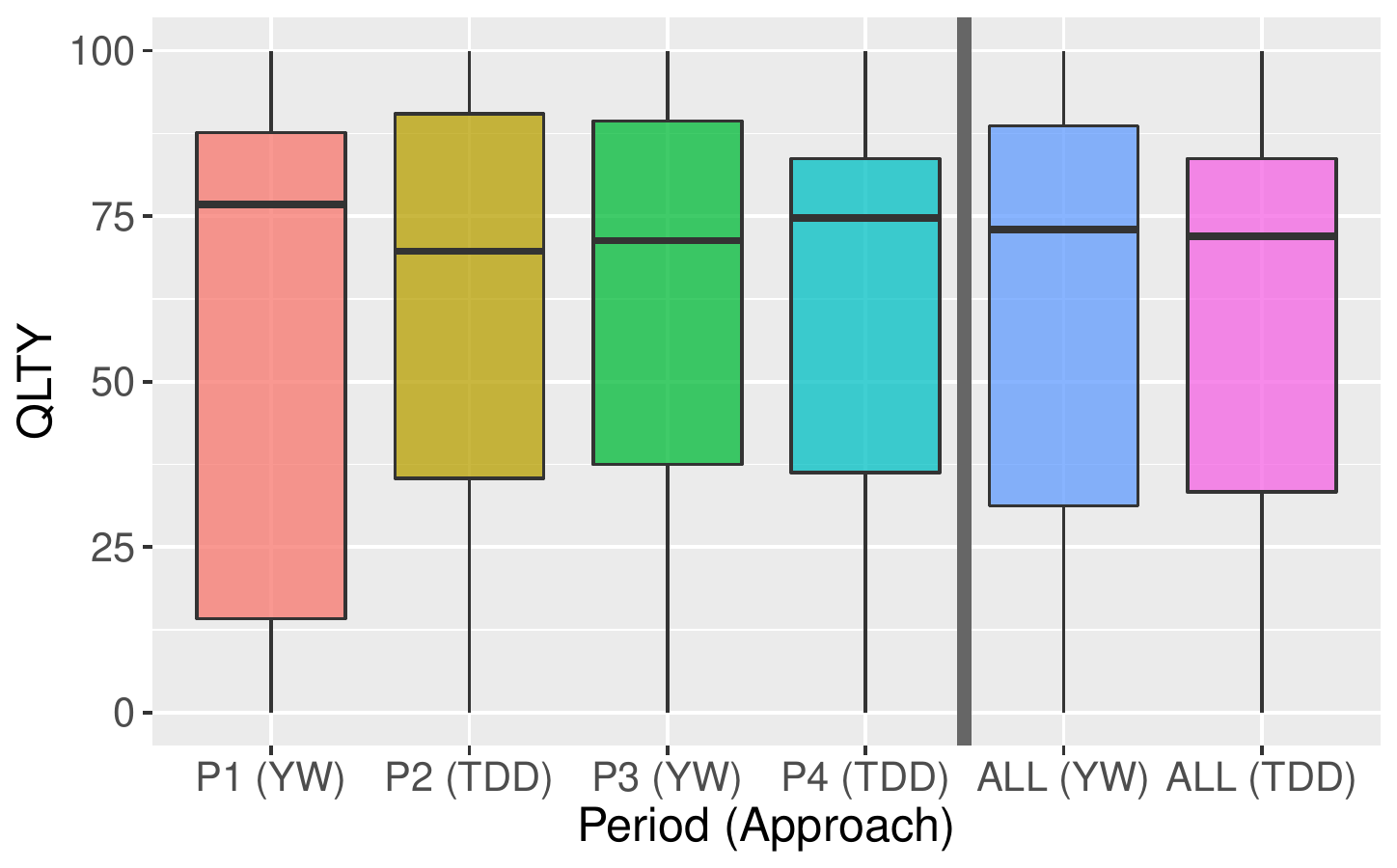}
\caption{Boxplots for QLTY grouped by Period and Approach.}\label{fig:boxplot:QLTY}
\end{figure}

In Figure~\ref{fig:boxplot:QLTY}, we report the boxplots for the \qlty variable grouped by Period or Approach. By looking at these boxplots, we can notice that there are not remarkable differences in the \qlty values when passing from one period to another one. In particular, if we compare the boxplots for P1 and P3---\ie same \yw treatment but different experimental objects---, we can notice that they overlap and the median level in P1 is higher than that in P3 (76.76 vs. 71.28 as shown in Table~\ref{tab:stats}).
As for the periods P2 and P4---\ie same \tdd treatment but different experimental objects---, the boxplots in Figure~\ref{fig:boxplot:QLTY} overlap and the median level is higher in P4 (69.72 vs. 74.76). That is, it seems that, when the experimental objects are \bsk and \mra (\ie in P1 and P4), the medians are higher. Therefore, the observed slight differences between P1 and P3, as well as between P2 and P4, seem to be due to the experimental objects. Summing up, there is no proof that \tdd is not retained with respect to \qlty. 

When comparing \tdd and \yw, the results (see Table~\ref{tab:stats} and Figure~\ref{fig:boxplot:QLTY}) do not suggest differences in the \qlty values (\eg on average, \qlty is equal to 60.81 for \tdd, while it is equal to 61.22 for \yw). This trend is confirmed when comparing P4 (\tdd) with P1 (\yw), as well as P2 (\tdd) with P3 (\yw),---\ie same experimental objects but different treatment. For instance, in P4 and P1, the mean values for \qlty are similar (58.53 vs. 59.39), although they applied either \tdd or \yw while tackling the same experimental objects. The comparison between P2 and P3 leads to a similar observation.

\subsubsection{\pro---Developers' Productivity}

\begin{figure}[t]
      \includegraphics[width=0.9\linewidth]{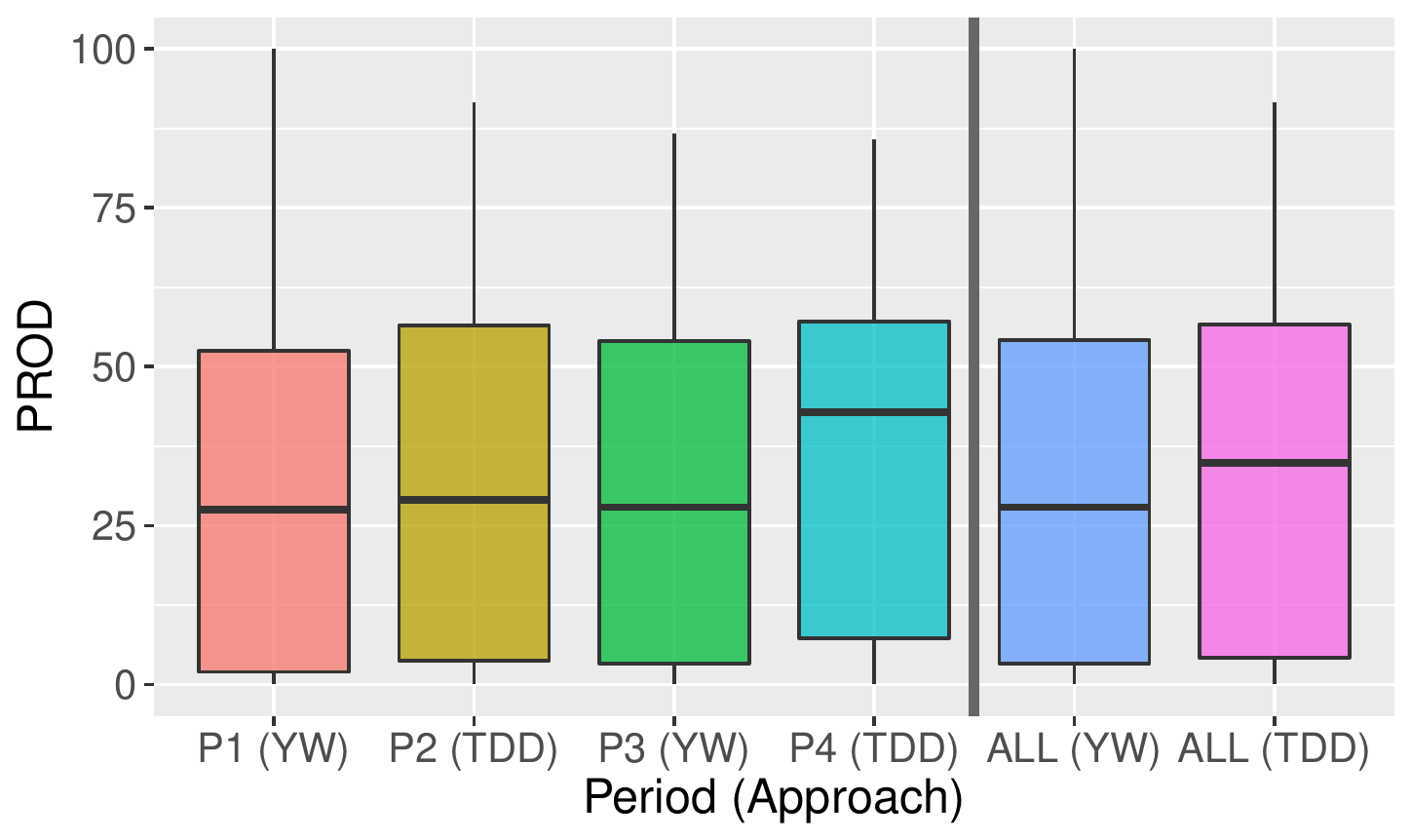}
\caption{Boxplots for PROD grouped by Period and Approach.}\label{fig:boxplot:PROD}
\end{figure}

By observing the boxplots for \pro in Figure~\ref{fig:boxplot:PROD}, we can notice that there is not a huge difference in the \pro values among the periods. Indeed, when comparing P2 with P4---same \tdd treatment but different experimental object---, we can observe that the boxplots overlap, although the median level for P4 is higher than for P2 (42.85 vs. 29.06). Such an improvement in the \pro values, when passing from P2 to P4, might be due to the \tdd retainment. As for the comparison between P1 and P3---same \yw treatment but different experimental object---, the boxplot for P1 is very similar to that for P3. That is, it seems that the \tdd knowledge the participants had in P3 did not affect~\pro.

Overall, it seems there is a slight difference in the \pro values between \tdd and \yw (see Table~\ref{tab:stats} and Figure~\ref{fig:boxplot:PROD}). This difference in favor of \tdd; for instance, the mean \pro value for \tdd is equal to 35.22, while that for \yw is equal to 32.55. By comparing pairs of periods in which the same experimental objects are used (but different treatments are applied), we can notice that the \pro values in P4 (\tdd) are better than those in P1 (\yw); \eg, the mean values are equal to 37.96 and 32.47 in P4 and P1, respectively. Namely, it seems that the participants who applied \tdd on \bsk and \mra achieved \pro values better than the participants who applied \yw on the same experimental objects. When comparing P2 (\tdd) and P3 (\yw)---the experimental objects were \gol and \ssh---,
it seems that there is no difference in the \pro values. For instance, the boxplots for P2 and P3 are very similar (see Figure~\ref{fig:boxplot:PROD}).

\subsubsection{\test---Number of Tests Written}
\begin{figure}[t]
      \includegraphics[width=0.9\linewidth]{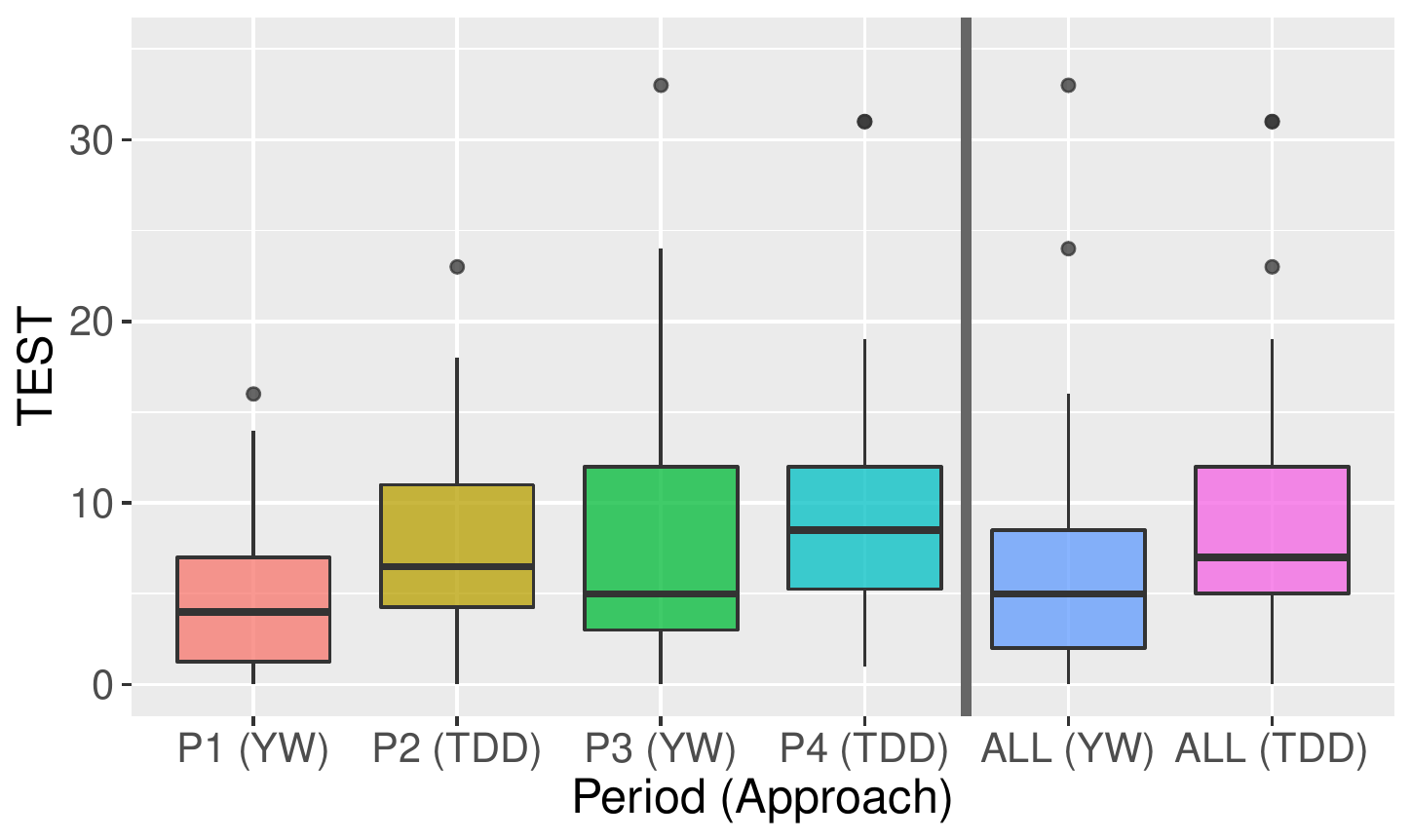}
\caption{Boxplots for TEST grouped by Period and Approach.}\label{fig:boxplot:TEST}
\end{figure}

The boxplots for \test are shown in Figure~\ref{fig:boxplot:TEST}. By observing them, we can notice differences in the \test values among the periods. In particular, if we compare the \yw treatments in P1 and P3, we can notice that the boxplot for P3 is higher than that for P1. The descriptive statistics reported in Table~\ref{tab:stats} confirm that the \test values are better in P3 than in P1---\eg the mean is equal to 7.93 for P3 and 4.93 for P1.
This difference might be due to the knowledge the participants had in P3 on \tdd. Namely, there might be a positive effect due to the \tdd retainment. On the other hand, when comparing the \tdd treatments in P2 and P4, the boxplots suggest a less pronounced difference in the \test values. Indeed, the boxplots for P2 and P4 overlap, even though the median level for P4 is higher than that for P2 (8.5 vs. 6.5). Summing up, the results suggest that \tdd can be retained with respect to \test.

As for the the comparison between \tdd and \yw, it seems that the participants who followed \tdd wrote more tests (see Table~\ref{tab:stats} and Figure~\ref{fig:boxplot:TEST}). For instance, they achieved, on average, \test values equal to 8.96 and 6.43 when following \tdd and \yw, respectively. If we consider only P1 and P4---same experimental object but different treatment---, we can observe a clear improvement in the \test values in P4; \eg the mean values are 4.93 and 10.1, respectively. Namely, the participants who applied \tdd in P4 seem to achieve higher \test values than those who applied \yw in P1 on the same experimental objects. Interestingly, the comparison between  P2 (\tdd) and P3 (\yw) does not highlight a remarkable difference. Namely, it seems that the distributions of the \test values for P2 (\tdd) and P3 (\yw) are quite similar (\eg the mean values are equal to 7.83 and 7.93, respectively), despite the application of either \tdd (in P2) or \yw (in P3) on the same experimental objects. This could indicate that the \tdd retainment had influenced the participants who followed \yw in P3.

\subsubsection{\mut---Fault-detection Capability of Tests Written}

\begin{figure}[t]
      \includegraphics[width=0.9\linewidth]{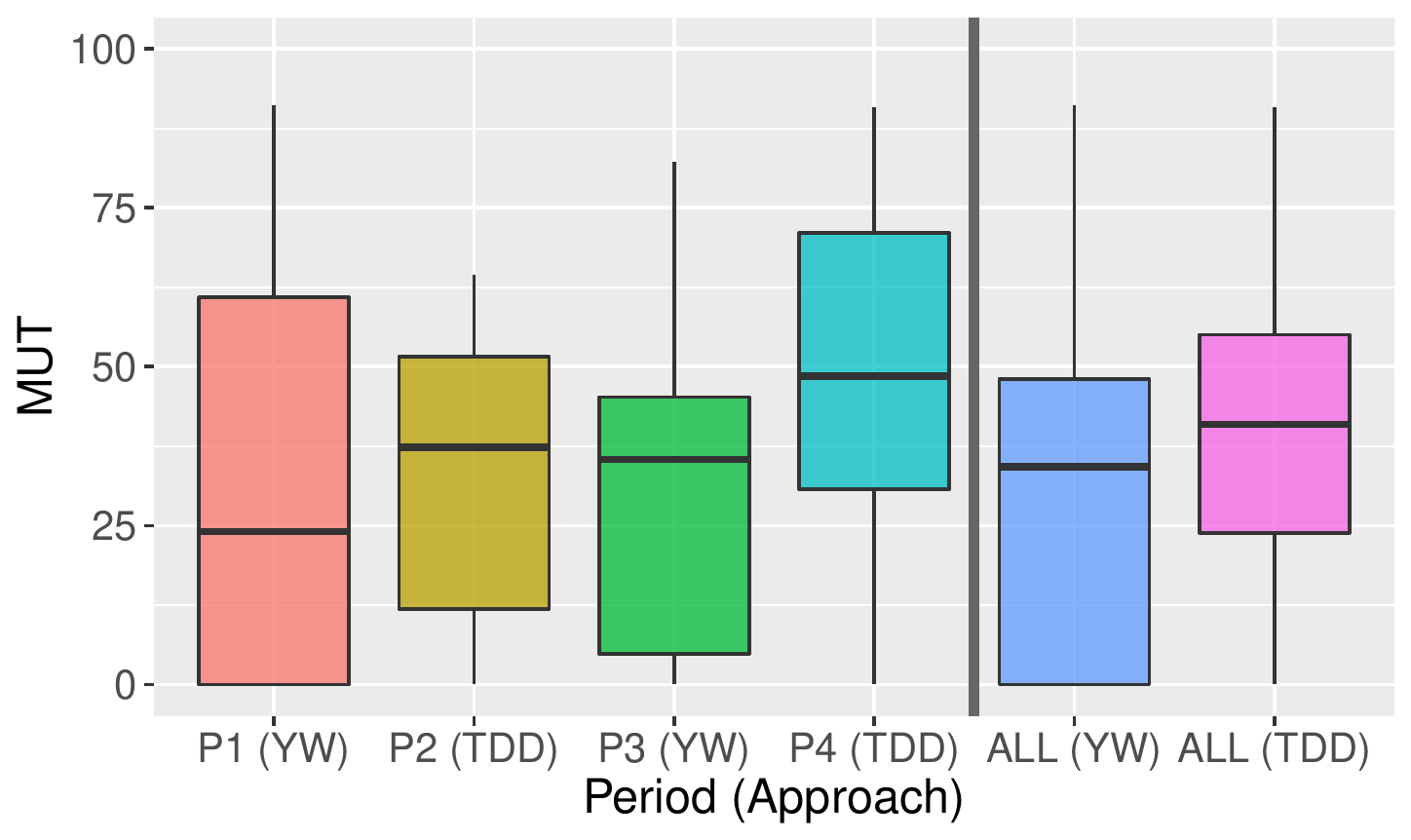}
\caption{Boxplots for \mut grouped by Period and Approach.}\label{fig:boxplot:MUT}
\end{figure}

By comparing the boxplots in Figure~\ref{fig:boxplot:MUT}, we can observe that there are differences in the \mut values among the periods. If we consider only P1 and P3---the periods in which \yw was applied---, the distributions of the \mut values look different. In particular, the boxplots for P1 and P3 overlap, but the latter is shorter and the median level is noticeably higher (24.1 vs. 35.43, see Table~\ref{tab:stats}). However, the mean values for P1 and P3 are almost identical (31.98 vs. 31.99). This is to say that there is a high variation in the \mut values in P1 that reduces in P3---after the participants knew \tdd. As for the comparison between the periods P2 and P4---those concerning \tdd---, we can observe two different distributions in Figure~\ref{fig:boxplot:MUT}. In particular, the distribution for P4 is noticeably higher than that for P2. That is, we can notice a clear improvement in P4 as far as the \mut values are concerned. The descriptive statistics in Table~\ref{tab:stats} remark this improvement (\eg the mean value passes from 32.07 in P2 to 48.52 in P4).

As for the overall boxplots of \tdd and \yw shown in Figure~\ref{fig:boxplot:MUT}, we can notice that the boxplot for \tdd is higher and shorter than that for \yw. Such a difference can be also observed by looking at Table~\ref{tab:stats}. For example, the mean values of \mut are 31.99 and 40.29 for \yw and \tdd, respectively. If we compare only P1 and P4---same experimental object but different treatment---, we can observe a clear improvement in the \mut values in P4; \eg the mean values are 31.98 and 48.52, respectively. As for the comparison between P2 and P3---same experimental object but different treatment---, the two distributions look similar; \eg the boxplots depicted in Figure~\ref{fig:boxplot:MUT} overlap. The descriptive statistics in Table~\ref{tab:stats} do not also highlight large differences in the \mut values between P2 and P3 (\eg the mean values are 32.07 and 31.99, respectively). Summing up, it seems that the differences in the \mut values reflect those in the \test values. The results from the inferential statistics could strengthen such a conclusion.

\subsubsection{\seq---Test-first Sequencing}

\begin{figure}[pt]
\subfloat[]{\label{fig:boxplot:SEQ}
      \includegraphics[width=0.45\linewidth]{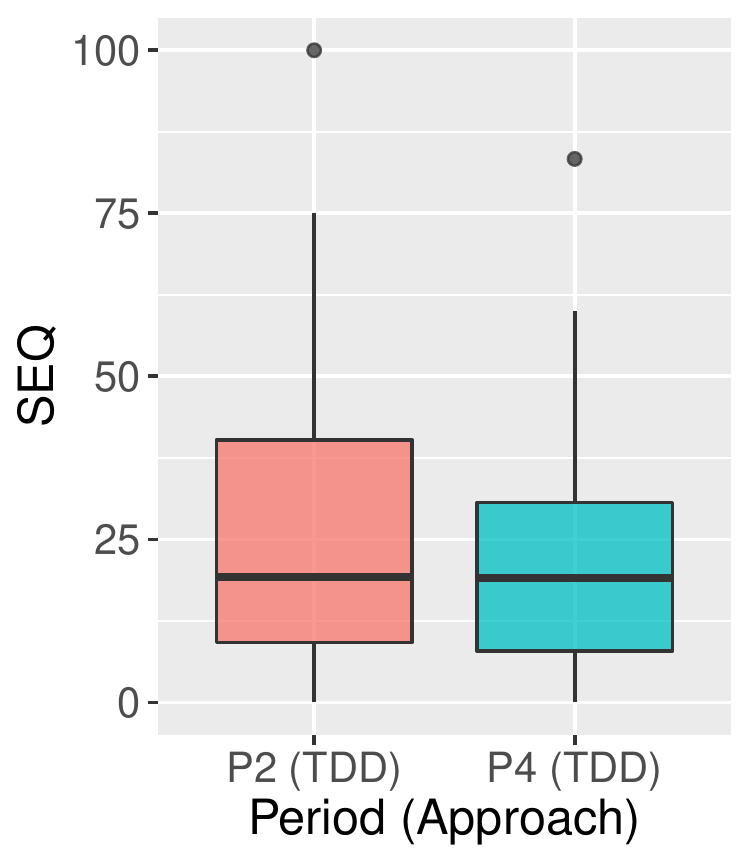}}
\subfloat[]{\label{fig:boxplot:GRA}
      \includegraphics[width=0.45\linewidth]{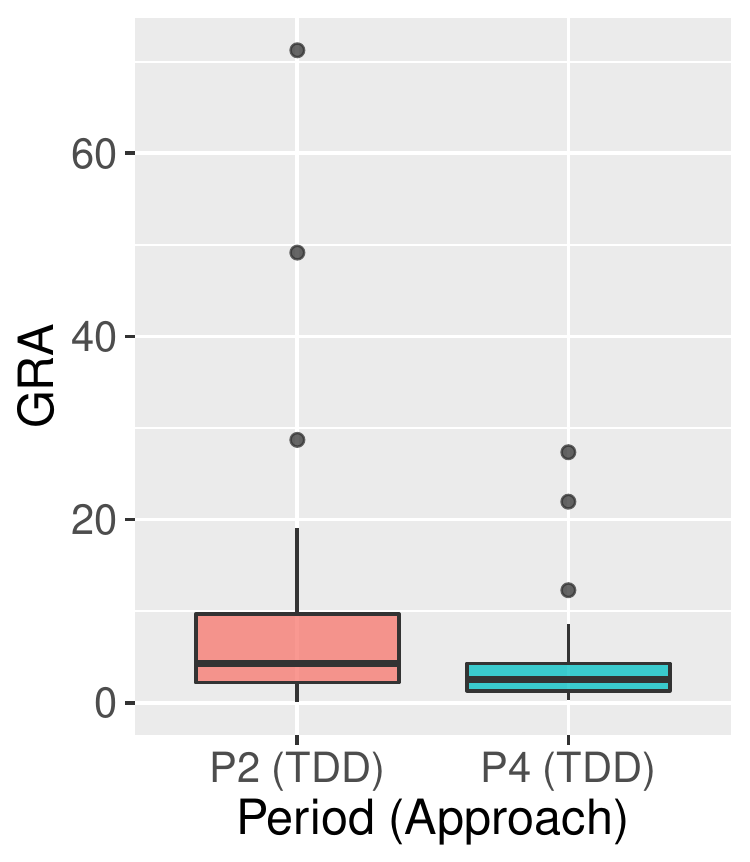}}

\subfloat[]{\label{fig:boxplot:UNI}
      \includegraphics[width=0.45\linewidth]{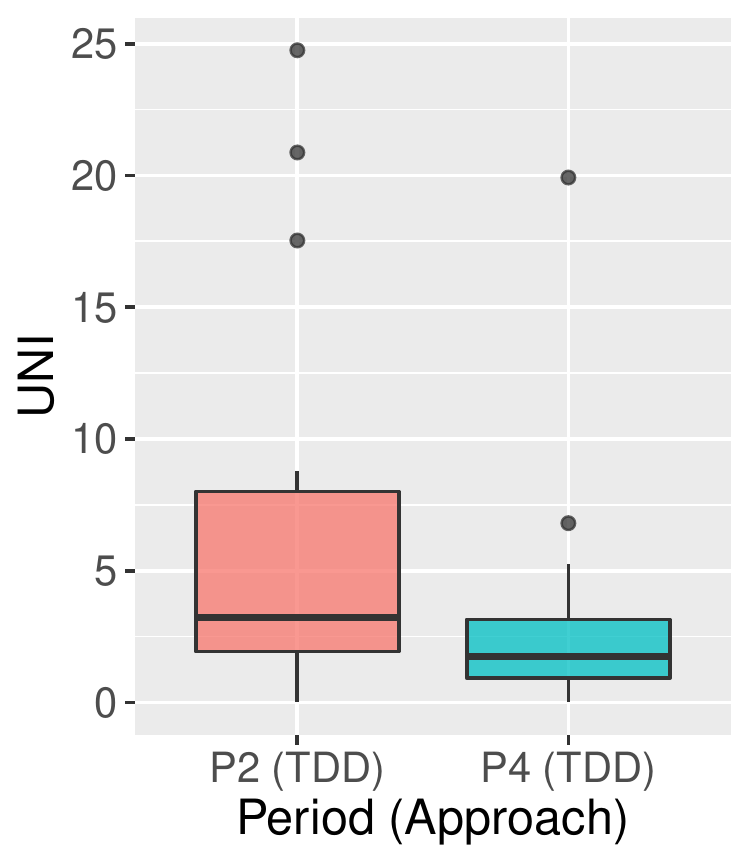}}
\subfloat[]{\label{fig:boxplot:REF}
      \includegraphics[width=0.45\linewidth]{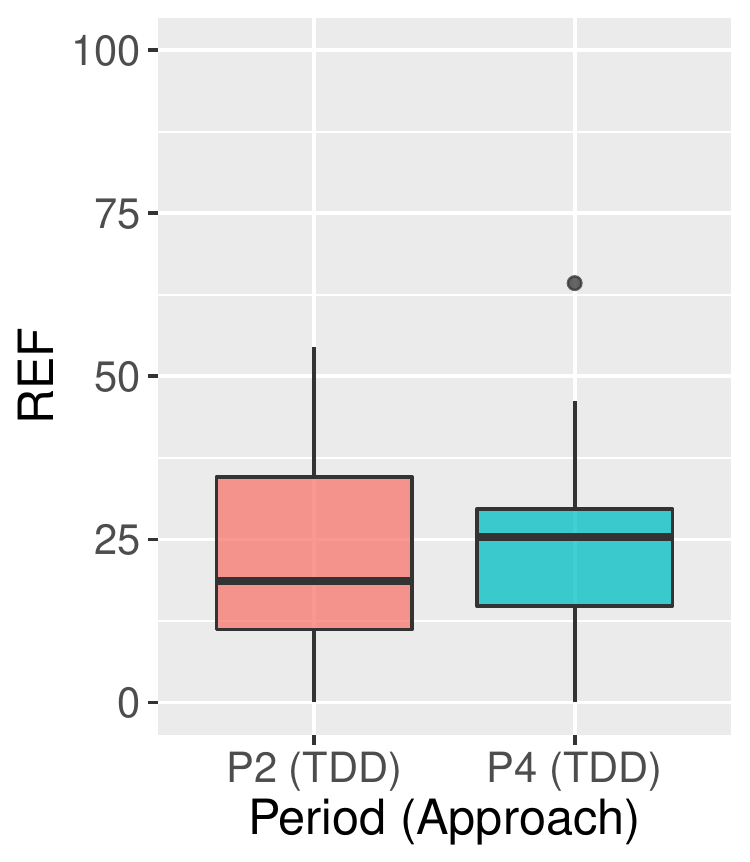}}

\caption{Boxplots for (a) \seq, (b) \gra, (c) \uni, and (d) \re grouped by Period (only P2 and P4).}\label{fig:boxplot:TDD}

\end{figure}

In Figure~\ref{fig:boxplot:SEQ}, we graphically summarize the distributions of the \seq values for P2 and P4---\ie the two periods in which the participants applied \tdd. By looking at this figure, we can observe that the boxplots for P2 and P4 overlap but the latter is shorter---\ie there is less variation in the \seq values in the second application of \tdd (as also confirmed by the SD values, 25.73 vs. 20.27, reported in Table~\ref{tab:stats}). Moreover, the median levels for P2 and P4 are very similar (19.21 vs. 19.09), even though, on average, the \seq values for P2 are higher than those for P4 (27.91 vs. 22.3). Summing up, it seems that there is no huge difference in the application of \tdd between P2 and P4 with respect to the \seq so suggesting a retainment of \tdd.

\subsubsection{\gra---Granularity}

The boxplots for \gra are depicted in Figure~\ref{fig:boxplot:GRA}. They seem to indicate a difference in the \gra values between P2 and P4. In particular, the boxplot for P4 is shorter and lower than that for P2---a lower \gra value is desirable (see Section~\ref{sec:dep:vars}). The descriptive statistics confirm this outcome; \eg the mean \gra values are equal to 4.68 and 10.29 for P4 and P2, respectively. That is, it seems that the participants retained \tdd when passing from P2 to P4 and due to the \tdd retainment there is an improvement in the \gra values in last~period.


\subsubsection{\uni---Uniformity}
The distributions for \uni depicted in Figure~\ref{fig:boxplot:UNI} look different. In particular, the boxplot for P2 is higher and larger than the one for P4. The descriptive statistics in Table~\ref{tab:stats} seem to also highlight differences in the \uni values; \eg the mean values are equal to 5.76 and 2.82 for P2 and P4, respectively. In other words, it seems that the participants achieved better \uni values in P4---a lower \uni value is desirable (see Section~\ref{sec:dep:vars})---and such an outcome could be due to a \tdd retainment.


\subsubsection{\re---Refactoring Effort}
Regarding the distributions for \re, summarized in Figure~\ref{fig:boxplot:REF}, we can observe that the distribution for P2 is quite similar to that for P4. Indeed, the boxplots for P2 and P4 overlap and the median level is higher in P4 (18.61 vs. 25.36). However, the \re values was, on average, similar: 22.89 in P2 and 23.69 in P4. Again, it seems the participants retained~\tdd with respect to~\re.

\subsection{Inferential Statistics}
In Table~\ref{tab:res:period}, we report the p-values from the LMM analysis methods. We highlight the effects that are statistically significant with the asterisk symbol.


\begin{table}[t]
\caption{Results (\ie p-values) from the inferential statistics. Note that LMM2\textsubscript{X} is built only when LMM1\textsubscript{X} indicates a statistically significant effect of Period. Moreover, LMM2\textsubscript{X} is not built for \seq, \gra, \uni, and \re.}
\label{tab:res:period}
\resizebox{\linewidth}{!}{
\begin{threeparttable}
\begin{tabular}{lllllll} \hline\noalign{\smallskip}
Variable & \multicolumn{3}{l}{LMM1\textsubscript{X}} & \multicolumn{3}{l}{LMM2\textsubscript{X}} \\
         & Period & Group & Period:Group & Approach & Group & Approach:Group \\ \noalign{\smallskip}\hline\noalign{\smallskip}
\qlty & 0.8837 & 0.6108 & $<$0.0001\tnote{$\ast$} & - & - & - \\
\pro  & 0.7973 & 0.8225 & $<$0.0001\tnote{$\ast$} & - & - & - \\
\test & 0.0002\tnote{$\ast$}  & 0.0617 & 0.4632 & 0.0005\tnote{$\ast$} & 0.0617 & 0.7706 \\
\mut & 0.0017\tnote{$\ast$} & 0.734 & $<$0.0001\tnote{$\ast$} & 0.0454\tnote{$\ast$} & 0.734 & 0.0025\tnote{$\ast$} \\
\seq & 0.4707 & 0.9864 & 0.5752 & - & - & -\\
\gra & 0.1992 & 0.2123 & 0.0581 & - & - & -\\
\uni & 0.021\tnote{$\ast$} & 0.4406 & 0.2211 & - & - & -\\
\re & 0.8581 & 0.5084 & 0.9611 & - & - & - \\
\noalign{\smallskip}\hline
\end{tabular}
\begin{tablenotes}
\item[$\ast$] \footnotesize Statistically significant effect.
\end{tablenotes}
\end{threeparttable}
}
\end{table}

\subsubsection{\qlty---External Quality of Implemented Solutions}
The assumptions of LMM1\textsubscript{\qlty} were both met. The residuals of the built LMM were normally distributed (the Shapiro test returned a p-value equal to 0.1166) and their mean was equal to zero.

The LMM analysis method do not allow us to reject HN1\textsubscript{\qlty} because the p-value for Period is 0.8837 (see Table~\ref{tab:res:period}), namely the effect of Period is not statistically significant. This means that there is neither a deterioration nor an improvement in the observed time span for the \qlty variable. 
LMM1\textsubscript{\qlty} also indicates a statistically significant interaction between Group and Period, which is due to the effect of the experimental objects (\eg regardless of the treatment, the distributions for \bsk are higher than those for \gol). As for the effect of Group, it is not statistically significant. We can therefore conclude that \tdd can be retained in terms of external quality of software~products.

Since LMM1\textsubscript{\qlty} does not indicate a statistically significant effect of Period, this implies that neither the effect of Approach is statistically significant (\ie there is no need to build LMM2\textsubscript{\qlty}). Therefore, we cannot reject HN2\textsubscript{\qlty}. This outcome seems to suggest that developing according to the \tdd approach does not influence the external quality of software products.


\subsubsection{\pro---Developers' Productivity}
The assumption of normality was not met for LMM1\textsubscript{\pro} since the Shapiro test returned a p-value equal to 0.0035. To meet this assumption, we needed a data transformation (square-root transformation, in particular). After transforming the data, the LMM assumptions were both satisfied---the residuals were normally distributed according to the Shapiro test (p-value=0.0524) and their mean was equal to zero.

As shown in Table~\ref{tab:res:period}, the effect of Period for LMM1\textsubscript{\pro} is not statistically significant (p-value=0.7973). Therefore, we cannot reject HN1\textsubscript{\pro}. This outcome indicates that the participants can retain \tdd as far as their productivity is concerned. LMM1\textsubscript{\pro} also includes a statistically significant effect, namely the one of Group:Period. Again, this is due to the effect of the experimental objects. Finally, there is no statistically significant effect of Group.

Since we cannot reject LMM1\textsubscript{\pro}, neither HN2\textsubscript{\pro} can be rejected. This said, it seems that developers who follow either the \tdd or \yw approach exhibit a similar productivity.

\subsubsection{\test---Number of Tests Written}
We needed a data transformation for the variable \test because the residuals for LMM1\textsubscript{\test} were not normally distributed (the p-value the Shapiro test returned was $<$0.0001). Therefore, we applied a log transformation, which allowed us to meet both LMM assumptions. In particular, the Shapiro test returned a p-value equal to 0.0797, suggesting that the residuals followed a normal distribution. The mean of the residuals was equal to zero.

LMM1\textsubscript{\test} shows a statistically significant effect of Period (p-value=0.0002). Therefore, we can reject HN1\textsubscript{\test} and conclude that Period significantly affects the number of tests the participants wrote. By looking at the boxplots in Figure~\ref{fig:boxplot:TEST}, we can notice that this statistically significant effect is not due to a deterioration of \tdd over time---the worst distribution of the \test values can be observed in P1 while the best distribution can be observed in P4. We can therefore conclude that developers following \tdd retain the ability to write unit tests. LMM1\textsubscript{\test} does not indicate other statistically significant~effects.

Since we found a statistically significant effect of Period, there can be a statistically significant effect of Approach. In this respect, the descriptive statistics and boxplots depict a clear difference in favor of \tdd when comparing P1 and P4---same experimental objects. This difference is also noticeable when comparing, in general, \tdd with \yw. To determine whether or not there is a statistically significant effect of Approach, we built LMM2\textsubscript{\test}. Since the the residuals of LMM2\textsubscript{\test} were not normally distributed (the p-value the Shapiro test returned was $<$0.0001), we performed a log-transformation for \test. Thanks to this transformation, the assumptions of LMM2\textsubscript{\test} were both met: the residuals followed a normal distribution (the p-value returned by the Shapiro test was 0.0562) and their mean was equal to zero.

As reported in Table~\ref{tab:res:period}, LMM2\textsubscript{\test} only includes a statistically significant effect, namely that of Approach (p-value=0.0005). Therefore, we can reject HN2\textsubscript{\test} and conclude that \tdd significantly and positively affects the number of tests the participants wrote.


\subsection{\mut---Fault-detection Capability of Tests Written}
The residuals of LMM1\textsubscript{\mut} were normally distributed (the Shapiro test returned a p-value equal to 0.5289) and their mean was equal to zero. Therefore, we did not need any data transformation for LMM1\textsubscript{\mut}.

LMM1\textsubscript{\mut} includes two statistically significant effects: one for Period (p-value=0.0017) and one for Period:Group (p-value$<$0.0001). The p-value of Period allows rejecting HN1\textsubscript{\mut}, thus recognizing that Period significantly affects the fault-detection capability of the written test suite. Moreover, we can observe an improvement of the \mut values in P4 with respect to any other period (\eg see Figure~\ref{fig:boxplot:MUT}). Therefore, we can conclude that developers can retain their ability to write tests in terms of both number of written tests and fault-detection capability of these tests. As for the p-value of Period:Group, it suggests that the fault-detection capability of the written tests can depend on the development task at hand.

Since we could reject HN1\textsubscript{\mut}, a statistically significant effect of Approach is possible. In this respect, LMM2\textsubscript{\mut} allows us to reject HN2\textsubscript{\mut} because the p-value of Approach is equal to 0.0454. That is, there is a statistically significant effect of Approach, in favor of \tdd, on the fault-detection capability of the written test cases. This outcome suggests that \tdd practitioners tend to write more tests than non-\tdd ones and, moreover, the fault-detection capability of these tests is significantly better. Again, the fault-detection capability of the written tests seems to depend on the development task at hand.

It is worth mentioning that, before applying LMM2\textsubscript{\mut}, we checked the LMM assumptions, which were both satisfied (\eg the Shapiro test returned a p-value equal to 0.247 indicating a normal distribution for the residuals).

\subsubsection{\seq---Test-first Sequencing}
The assumption of normality was not satisfied for LMM1\textsubscript{\seq} (the Shapiro test returned a p-value equal to 0.0005). To satisfy both LMM assumptions, we square-transformed the \seq variable. After this data transformation, the residuals were normally distributed according to the Shapiro test (p-value=0.3846) and their mean was equal to zero.

By looking at the p-values in Table~\ref{tab:res:period}, we can notice that LMM1\textsubscript{\seq} do not include a statistically significant effect for Period. This outcome suggests that developers can retain \tdd  with respect to the test-first sequencing. As for the other p-values, they indicate a statistically significant effect for neither Group nor Period:Group. The p-value for Period:Group seems to indicate that developers' ability of following the test-first dynamic is not affected by the development task (\ie the experimental~object).

\subsubsection{\gra---Granularity}
To met the assumptions of LMM1\textsubscript{\gra}, we had to apply a log-transformation. This is because the Shapiro test indicated a violation of the normality assumption of the residuals (p-value$<$0.0001). Thanks to the log-transformation, we satisfied the assumption of normality of the residuals (the p-value returned by the Shapiro test was 0.0568) as well as that concerning their mean.

As shown in Table~\ref{tab:res:period}, the effect of Period for LMM1\textsubscript{\gra} is not statistically significant although the boxplots in Figure~\ref{fig:boxplot:GRA} highlighted an improvement in P4 (with respect to P2). On average, the granularity of the development cycles was 4.26 and 2.5 minutes. These outcomes suggest that developers can retain their ability of following short cycles when applying the \tdd approach. The p-values for Group and Period:Group did not highlight any statistically significant difference. Similarly to the test-first sequencing characteristic, it seems that the granularity of development cycles is not affected by the tasks.

\subsubsection{\uni---Uniformity}
We applied a log-transformation to meet the assumptions of LMM1\textsubscript{\uni}. This is because the residuals were not normally distributed according to the Shapiro test (p-value$<$0.0001). By applying the log-transformation we met both LMM assumptions (in particular, the Shapiro test returned a p-value equal to~0.065).

The p-values in Table~\ref{tab:res:period} indicate a statistically significant effect of Period (0.021). The distributions of the \uni values (\eg see Figure~\ref{fig:boxplot:UNI}) suggests that, in P4, the development cycles of the participants who applied \tdd were more uniform as compared to P2. This is to say that developers can retain the \tdd characteristic of carrying out uniform development cycles. Finally, there is a statistically significant effect for neither Group nor Period:Group---\ie it seems that the tasks do not influence the uniformity of development cycles.

\subsubsection{\re---Refactoring Effort}
We did not need any data transformation for LMM1\textsubscript{\re} because the assumptions were both met. In particular, the residuals followed a normal distribution (the Shapiro test returned a p-value equal to 0.1134) and their mean was equal to zero.

The results, shown in Table~\ref{tab:res:period}, do not highlight any statistically significant effect. Concluding, it seems that the refactoring effort is retained when practicing \tdd. Again, the tasks seem to not influence the refactoring effort.

\section{Discussion}\label{sec:discussion}

In this section, we first answer the RQs to delineate the main findings of our cohort study. We then discuss these findings and present their practical implications. Finally, we discuss the threats that may have affected the validity of these findings.

\subsection{Answering Research Questions}
We observed no deterioration, during the considered time span, in the external quality of the solutions our participants implemented (\ie \qlty), their productivity (\ie \pro), and number of tests they wrote (\ie \test). Furthermore, the way in which the participants followed the process underlying \tdd (\ie in terms of \seq, \gra, \uni, and \re) did not deteriorate over time. On the other hand, we observed a significant improvement in the number of tests, which led to a better fault-detection capability of these tests (\ie \mut), after the participants had known and practiced \tdd. The uniformity of the cycles enhanced with time as well. On the basis of these results, we can answer RQ1 as follows.

\begin{mdframed}[backgroundcolor=gray!10]
\noindent
\textbf{Developers retain \tdd over about six months. In particular, while the external quality of software products and developers' productivity are neither deteriorated nor improved over that time span, the amount of written test increases as well as their fault-detection capability. Moreover, the way in which developers follow \tdd remains constant in the considered time span, except for the uniformity of the process underlying \tdd, which is more uniform over time.}
\end{mdframed}
This outcome is perhaps not overly surprising, but evidence needs to be obtained through empirical studies to move from opinions and common sense to facts (\eg~\cite{kitchenham02preliminary,basili99}), as well as to have a first understanding of \tdd retainment on several constructs.

As for the comparison between \tdd and \yw, we observed differences in favor of the former only when considering the amount of written tests (\ie \test). Such a difference is also present when considering the fault-detection capability of the tests written (\ie \mut). Accordingly, we can answer RQ2 as~follows.

\begin{mdframed}[backgroundcolor=gray!10]
\noindent
\textbf{While \tdd does not increase (or decrease) the external quality of software products and developers' productivity, it leads developers to create larger test suites with a higher fault-detection capability.}
\end{mdframed}


\subsection{Overall Discussion and Future Research}
We show that developers retain \tdd over a time span of about six months. This finding is in line with the preliminary empirical evidence gathered by Latorre~\cite{Lat14} on the retainment of \tdd---where he observed that, in a six-month time span, three developers retained \tdd in terms of developers' performance and conformance to \tdd. Based on the current empirical evidence on \tdd, we can deduce that the investment in training new \tdd practitioners is not squandered---it is preserved at least over a time span of six months. Furthermore, previous work has also shown that developers can correctly apply \tdd after a short practical session only~\cite{Lat14}. Accordingly, we can postulate that the investment in training new \tdd practitioners is reasonable. The question that now arises is how long such an investment is preserved, \ie how long developers retain \tdd. To answer this question, further longitudinal cohort studies are needed.  Our study has, therefore, the merit to increase the body of knowledge on the retainment of \tdd as well as to delineate new possible investigations on how long developers retain \tdd.

Among the investigated constructs, we observed that the retainment of \tdd is particularly noticeable in the amount of tests written since it increased after the participants had known and practiced \tdd. This seems to suggest that \tdd raises developers' awareness about the importance of writing unit tests; furthermore, these tests exhibit a higher fault-detection capability. Therefore, we advise instructors to teach \tdd when training new unit testers.

As compared to those who follow a non-\tdd development process, we also observe that developers practicing \tdd, write more unit tests that have a better fault-detection capability as well.
This finding is in line with that by Erdogmus \etal~\cite{Erdogmus:2005} so bringing further evidence that \tdd has a positive effect on the number of written tests. Again, our findings go in the direction of increasing the body of knowledge on the effect of \tdd.

Having more unit tests, with a higher fault-detection capability, should encourage software companies that value unit testing (\eg in order to create regression test suites for continuous integration) to adopt \tdd. Possible benefits deriving from having many tests with high fault-detection capability could be early fault detection and facilitated comprehension of unfamiliar source~code (\eg it has been showed that developers dealing with an unfamiliar code base look for examples of input/output values to better understand that code base~\cite{Sillito:2008}---unit tests contain such a kind of examples).

Our results do not highlight any improvement due to \tdd, with respect to the external quality of software products and developers' productivity, so contributing to the null results in \tdd research (\eg \cite{Fucci:2016,fucci2013replicated}). However, unlike previous studies, we observe that \tdd has no effect even when the same individuals are tested again several months later, under similar conditions. Time did not reduce novice developers' performance when TDD was applied, hinting at the fact that they soon regained familiarity with this technique, similarly to what Latorre reported for the junior developers involved in his study~\cite{Lat14}. Although carrying out cohort longitudinal studies---in particular, with several observations over a long time span---is difficult in Software Engineering (\eg controlling for maturation or keeping motivated the participants), we put forward the idea that we might not be looking long enough (rather than hard enough) for the claimed effects of \tdd to become apparent. As a starting point towards this direction, we recommend longitudinal studies in academia capable of following students' careers over several years and thus achieving a good amount of control (\eg based on grades). We advise this kind of investigation very risky and difficult to conduct. However, our study seem to justify future research on this matter.


We observe neither a deterioration nor an improvement over time in the external quality of software products and developers' productivity. A possible cause for this finding is that, with the only exception of the uniformity of the process underlying \tdd, the way in which the participants followed \tdd (\ie test-first sequencing, granularity, and refactoring effort) remained constant in the considered time span. Past work has shown that (external) quality and productivity improvements are primarily positively associated with the granularity and uniformity of the process underlying \tdd~\cite{Fucci:2017}. Therefore, it is possible that observing a significant difference in the uniformity of the process underlying \tdd is not enough to show alone a significant difference in the external quality of software products and developers' productivity.


Finally, to bring further evidence on both retainment of \tdd and its effects, we foster replications of our longitudinal cohort study. To this end, we made available on the web our laboratory package:\begin{itemize}
    \item \url{https://doi.org/10.6084/m9.figshare.6850013.v1}.\footnote{This is the laboratory package of Fucci \etal's study~\cite{Fucci:2018:ESEM}---awarded as ``Open Data Recognition'' at ESEM 2018. In case of acceptance of the paper, we are going to update that laboratory package with the new data we gathered (\eg data on fault-detection capability of written tests), which are temporarily available on: \url{https://tinyurl.com/Raw-Data-Ext-ESEM}.}
\end{itemize}

\subsection{Threats to Validity}
To determine the threats that could affect the validity of our study, as well as its results, we followed the guidelines by Wohlin \etal~\cite{Wohlin:2012}. Despite our effort to lessen or avoid as many threats as possible, some of them are unavoidable. This is because reducing or avoiding a kind of threat (\eg internal validity) may intensify or introduce another kind of threat (\eg external validity)~\cite{Wohlin:2012}. Since we conducted the first cohort study investigating the theory of \tdd retainment, we preferred to reduce threats to internal validity (\ie make sure that the cause-effect relationships were correctly identified), rather than being in favor of external~validity.

\subsubsection{Threats to Internal Validity} \label{sec:internalValidity} This kind of threat concerns internal factors of our study that could have affected the results.

\begin{itemize}
    \item \textbf{Selection.} The participants in our study were volunteers. This might threaten the validity of the results because volunteers might be  more motivated than the overall population~\cite{Wohlin:2012}.
    \item \textbf{Diffusion or treatments imitations.} To prevent that participants exchanged information during the development tasks, at least two researchers monitored them. Moreover, the participants were assigned to each workstation in the laboratory alternating the experimental objects. We also prevented the diffusion of experimental materials by gathering them at the end of each task and asking the participants not to talk with their classmates about the tasks they had implemented. Despite our effort to lessen this threat, we cannot exclude its presence, \eg some participates might have exchanged information about the tasks outside the laboratory.
    \item \textbf{Resentful demoralization.} Some participants  might not perform as good as they generally would since they might have received a less desirable treatment (or tasks). If this threat had existed in our study, it would have equally affected \tdd and \yw.
    \item \textbf{Maturation.} The control over participants was checked by making sure that the students attended the same courses between the first observation and the last one. This might affect the obtained results. Moreover, it seems that four participants did not launch Besouro at the beginning of the \tdd sessions. To have all data paired, we did not take into account these participants in the analyses of \seq, \gra, \uni, and \re. If the participants had launched Besouro, the results might have changed.
\end{itemize}

\subsubsection{Threats to Construct Validity} They concern the relationship between theory and observation.
\begin{itemize}
    \item \textbf{Mono-method bias.} We used a single measure for each investigated construct. This might affect the validity of the results if there was a measurement bias. To mitigate this threat, we used well-known measures~\cite{Fucci:2017,Tosun:2017,Fucci:2016,Erdogmus:2005}.
    \item \textbf{Hypotheses guessing.} Participants in  an empirical study might guess the study goals and then behave according to their guesses. Although we did not disclose our study goals to the participants (\ie we did not tell them that they were involved in an empirical study, nor how it was planned and how assignments were distributed among participants), someone might have guessed the goals and changed their behavior accordingly.
    \item \textbf{Evaluation apprehension.} Some people are afraid of being evaluated. To mitigate this threat, we informed the participants that they would not be evaluated on the basis of their performance in the study.
    \item \textbf{Restricted generalizability across constructs.} We found that \tdd positively affects the number of tests written and that there is retention of \tdd on the investigated constructs. However, we cannot exclude that \tdd has some side effects that our study was not able to reveal, as well as that \tdd is not retained for non-investigated constructs. To deal with this threat, we selected the dependent variables according to industrial needs~\cite{CSP11} as well as our previous experiences~\cite{Fucci:2017,Fucci:2016,FTJ15}.
\end{itemize}

\subsubsection{Threats to Conclusion Validity} This kind of threat concerns the relationship between dependent and independent variables.
\begin{itemize}
    \item \textbf{Reliability of treatment implementation.} Some participants might have followed the \tdd approach more strictly than others. This could threaten the validity of the obtained results. Moreover, some participants might have followed the \tdd approach when they were asked to use the \yw approach (\ie in P3) or vice-versa (\ie in P2 and P4). To mitigate this threat, we reminded the participants several times to follow the treatment they were assigned~to. Another threat concern the time span considered in our study (\eg a longer time span could negatively affect the \tdd retainment). Given that there is no guideline on the time-span duration and a previous study has considered a six-month time span~\cite{Lat14}, we believed that a time span of about six months sufficed to (preliminary) study the \tdd retainment even though we advise future research on the effect of longer time spans on the \tdd retainment. However, a larger time span would introduce some problems to counteract: participants could drop out (shrinking the sample size and decreasing the amount of data collected), while others could simply loose the motivation to participate.

    \item \textbf{Random heterogeneity of participants.} There is always heterogeneity in a study group~\cite{Wohlin:2012}. To lessen this threat, our study group consisted of students with similar background---\ie students taking the same courses in the same university with similar development experience. We collected the general information of the sample through a questionnaire before assigning students to the groups.
    \item \textbf{Reliability of measures.} To measure sequencing, granularity, uniformity, and refactoring effort, we exploited the Besouro plugin. Its use might threaten the validity of our results. However, Besouro represents the state-of-the-art tool for capturing the cycles when applying the TDD approach (\eg~\cite{Fucci:2017,Romano:2016,FTJ15,Romano:2017}).

\end{itemize}

\subsubsection{Threats to External Validity} External validity threats concern the ability to generalize the results.
\begin{itemize}
    \item \textbf{Interaction of selection and treatment.} The participants of our cohort study were students and this might affect the generalizability of findings with respect to software professionals.
    However, the used development tasks did not require a high level of professional programming experience. Therefore, we believe that involving students in our study could be considered appropriate, as suggested in the literature~\cite{Carver:2003,Host:2000}. Moreover, the use of students as participants bring also a number of advantages like having a homogeneous group of participants, having an opportunity of obtaining preliminary empirical evidence, \etc~\cite{Carver:2003}. In this respect, thanks to the use of students as participants, we could bring some evidence on the \tdd retainment through a longitudinal cohort study.

    \item \textbf{Interaction of setting and treatment.} The used code katas might not be representative of real-world development tasks. However, code katas are widely utilized to assess \tdd (\eg~\cite{Fucci:2017,Tosun:2017,Fucci:2016,Erdogmus:2005}) because they allows having a better control over the participants. For example, such code katas are conceived to be completed in an experimental session of approximately three hours.

\end{itemize}

\section{Conclusion}\label{sec:conclusion}
In this paper, we present the results from a quantitative longitudinal cohort study with 30 novice developers to investigate: \textit{(i)}~the retainment of \tdd over a time span of about six months and \textit{{(ii)}}~the effects of \tdd. We found that developers retain \tdd over the considered time span. This empirical evidence on the retainment of \tdd, together with past preliminary empirical evidence, allows us to conclude that the investment to train \tdd developers is guaranteed at least for a time span of six months. We also show that \tdd has no effect on external quality of software products and developers' productivity even when novice developers are tested again about six months later, under similar conditions. However, we observed that the participants practicing \tdd wrote significantly more unit tests, with a better fault-detection capability, than those practicing a non-\tdd approach. These results should foster software companies that value unit testing to have teams of developers that know \tdd.


\section*{Acknowledgements}
We would like the thank the students for their participation in our study.


\bibliography{mybibfile}

\end{document}